\documentclass[9pt,twocolumn,twoside]{osajnl}

\journal{josab} 

\setboolean{shortarticle}{false}

\usepackage[T1]{fontenc}
\usepackage{array}
\usepackage{multirow}
\usepackage{amsmath}
\usepackage{amssymb}
\usepackage{graphicx, epstopdf}
\usepackage{hyperref}
\usepackage{blindtext}
\usepackage{caption}
\usepackage{cleveref}
\usepackage{comment}

\makeatletter

\pdfpageheight\paperheight
\pdfpagewidth\paperwidth

\providecommand{\tabularnewline}{\\}

\makeatother

\usepackage{babel}
\begin{document}

\title{Theory of Four-Wave Mixing of\\
Cylindrical Vector Beams in Optical Fibers}

\author[1]{E. Scott Goudreau}
\author[1,2,3]{Connor Kupchak}
\author[1,2,3,4]{Benjamin J. Sussman}
\author[1,5]{Robert W. Boyd}
\author[1,2,3,*]{Jeff S. Lundeen}

\affil[1]{Department of Physics, Centre for Research in Photonics, University
of Ottawa, 25 Templeton St, Ottawa, ON, K1N 6N5 Canada}
\affil[2]{Department of Electronics, Carleton University, 1125 Colonel By Dr, Ottawa, ON, K1S 5B6 Canada}
\affil[3]{Joint Centre for Extreme Photonics, Ottawa, ON, Canada}
\affil[4]{National Research Council of Canada, 100 Sussex Dr, Ottawa, ON, K1N 0R6 Canada}
\affil[5]{The Institute of Optics and Department of Physics and Astronomy,
University of Rochester, Rochester NY 14627, USA}

\affil[*]{Corresponding author: jlundeen@uottawa.ca}

\begin{abstract}
Cylindrical vector (CV) beams are a set of transverse spatial modes
that exhibit a cylindrically symmetric intensity profile and a variable
polarization about the beam axis. They are composed of a non-separable
superposition of orbital and spin angular momentum. Critically, CV
beams are also the eigenmodes of optical fiber and, as such, are of
wide-spread practical importance in photonics and have the potential
to increase communications bandwidth through spatial multiplexing.
Here, we derive the coupled amplitude equations that describe the
four-wave mixing (FWM) of CV beams in optical fibers. These equations
allow us to determine the selection rules that govern the interconversion
of CV modes in FWM processes. With these selection rules, we show
that FWM conserves the total angular momentum, the sum of orbital
and spin angular momentum, in the conversion of two input photons
to two output photons. When applied to spontaneous four-wave mixing,
the selection rules show that photon pairs can be generated in CV
modes directly and can be entangled in those modes. Such quantum states
of light in CV modes could benefit technologies such as quantum key
distribution with satellites. 
\end{abstract}
\maketitle
\section{Introduction}

The term structured light refers to optical beams whose intensity,
phase, or polarization are non-uniform across the beam's transverse
profile. Cylindrical vector (CV) beams are a type of structured light
that exhibits intensity and polarization profiles that are spatially-dependent
but also exhibit symmetry under discrete rotations about the beam
axis. Specifically, the modes for such structured light beams, CV
modes, are described by a superposition of product states between
the spin angular momentum (SAM) and orbital angular momentum (OAM)
degrees of freedom. Utilizing the degrees of freedom available in
the transverse profile of an optical beam can benefit many applications.
These can include more complex mode-division multiplexing techniques
to increase communications bandwidth. Such multiplexing techniques
have already been demonstrated with scalar OAM modes \citep{Bozinovic20131545,Liu2018}
and vector beam modes \citep{Yuan2017, Li2018}. Similarly from a
quantum optics viewpoint, CV modes can increase the information capacity
available in single photons via the additional OAM and SAM degrees
of freedom. In free-space quantum key distribution, the rotational
symmetry of CV modes can be exploited to alleviate the need for rotational
alignment between sending and receiving parties \citep{Vallone2014}.
Outside of communications, the behavior of optical beams in CV modes
are of fundamental interest. For example, radially polarized modes
focus to smaller spot sizes than Gaussian modes of a comparable beam
waist \citep{PhysRevLett.91.233901}, and azimuthally polarized modes
produce a longitudinal magnetic field at their focus.

Thorough understanding of nonlinear optical processes is vital to
many practical applications. An example is telecommunications, where
unwanted nonlinear interactions between optical pulses in fiber present
a roadblock for increasing signal power, and thus the bandwidth \citep{Mitra2001}.
To date, the theory for four-wave mixing (FWM) in fiber systems has
been developed for both uniformly polarized light \citep{Lin2004,Garay2016}
and spatial modes \citep{Agrawal2007,Pourbeyram2016,Nazemosadat2016}.
Despite their potential for increasing communication bandwidth, the
nonlinear optics of structured light and vector modes is in its infancy
\citep{Torres2016,Saaltink201624495,Arlt19992438}. In order to address
this, here we derive the coupled amplitude equations that describe
four-wave mixing of CV and other complex modes. Using this, we derive
a set of general selection rules for the allowed mixing processes
between CV modes. FWM processes can convert photons between different
CV modes and may provide insight into conversion processes that involve
structured light and the conservation of angular momentum of light.
Moreover, these FWM transitions can potentially produce mode-entangled
CV photons through spontaneous four-wave mixing (SFWM) photon pair
generation.

\section{Optical Mode Descriptions}

\begin{figure}
\protect\includegraphics[width=0.9\columnwidth]{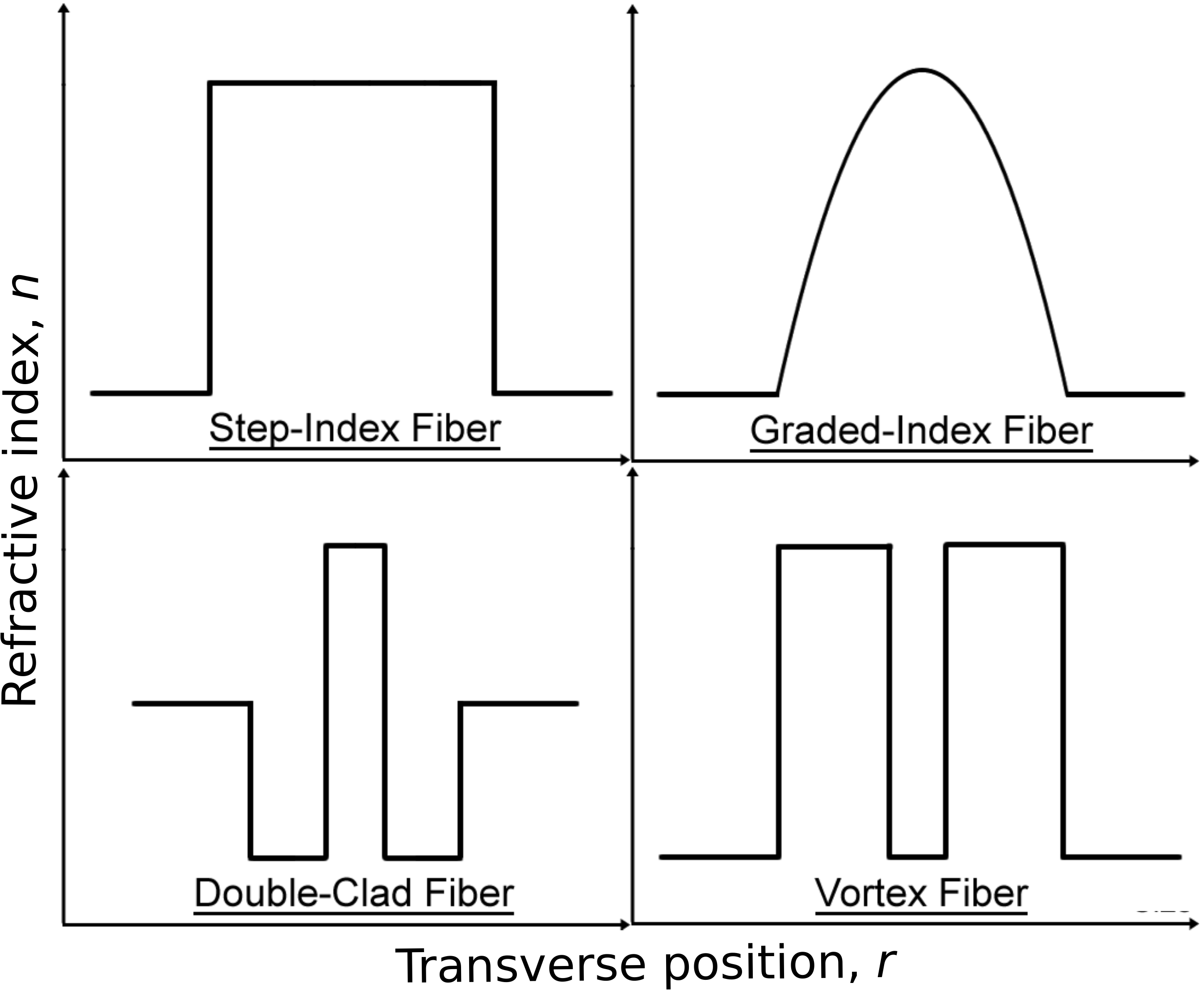}

\protect\caption{Four examples (step-index, graded-index, double-clad, and vortex)
of fiber types which support CV modes that are directly applicable
to this work.\label{fig:Four-examples}}
\end{figure}

We begin with a review of optical spatial modes that includes a succinct
general mathematical description that is suitable for deriving selection
rules. Then, starting with the nonlinear optical wave equation, we
derive general coupled-amplitude equations for four-wave mixing of
fields that vary spatially in intensity, phase, and polarization.
The derived FWM theory applies to both free-space environments such
as bulk nonlinear media and to weakly-guiding cylindrically symmetric
fibers (examples are shown in Fig. \ref{fig:Four-examples}). We end
by focusing on examples of these fields, particularly CV modes, but
also circularly polarized OAM modes, and modes that are eigenstates
of the total angular momentum along $z$, the beam or fiber axis.
For these examples, we derive and present selection rules. These spatial
mode solutions are highly relevant in the development of photonic
devices since CV modes are eigenmodes of all weakly-guiding cylindrically-symmetric
waveguides, such as standard optical fibers \citep{SnyderLove:1983,Chen2017}.
Furthermore, cylindrical vector modes represent approximate solutions
to the paraxial vector wave equation and, in free-space, follow the
intensity distribution of Laguerre-Gauss modes \citep{Zh2009}.

\begin{figure}[ht]
\protect\includegraphics[width=0.9\columnwidth]{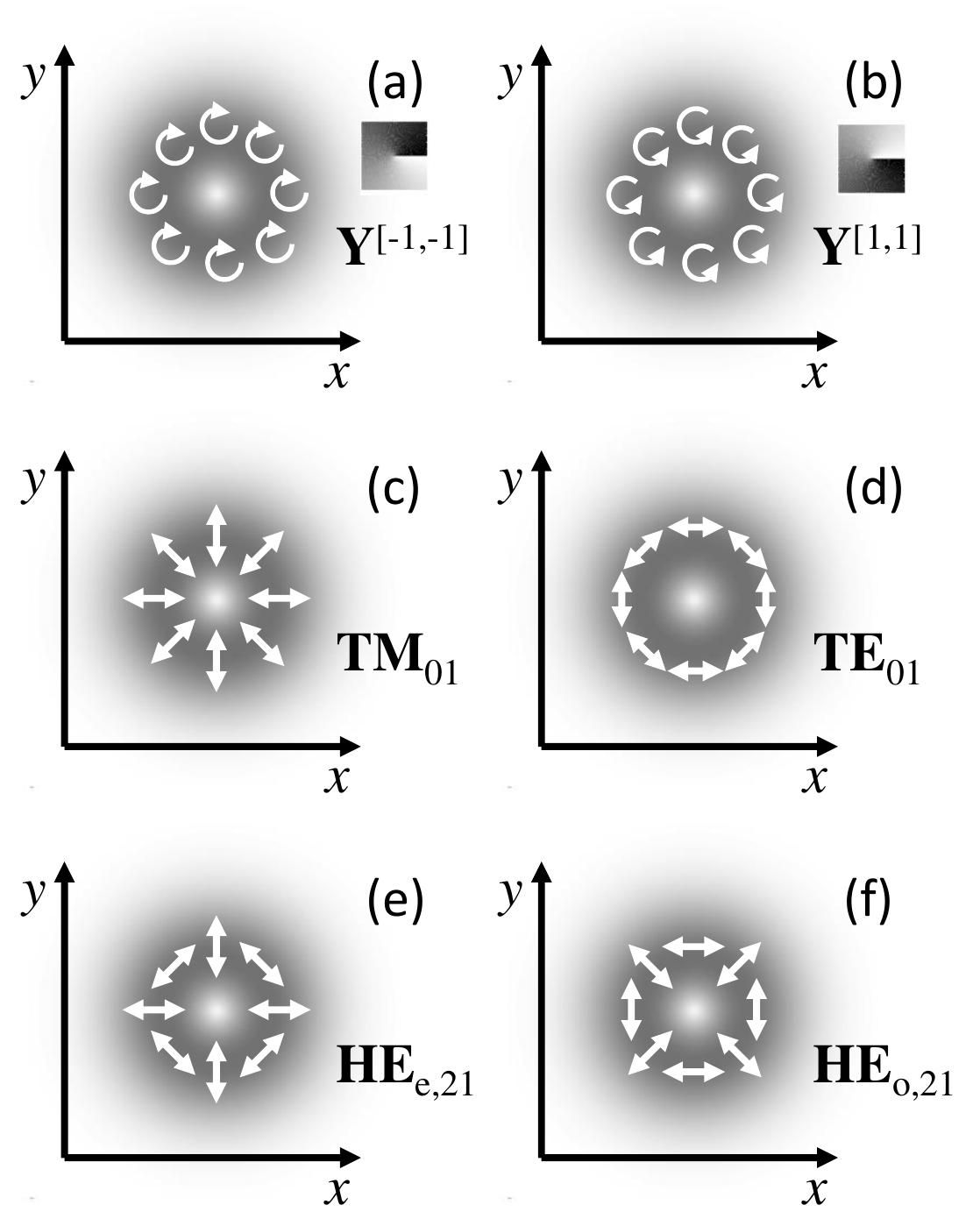}\protect\caption{Structured light modes.  The grey-scale gives the intensity, the arrows indicate the polarization, and phase is given by the grey-scale inset. All the depicted modes have the same orbital angular magnitude, $|l|=|L|=1$. The top two plots are spin and orbital angular momentum eigenstate
modes $\mathrm{\mathbf{Y}}$ with (a) $s=1$ and $l=1$ and (b) $s=-1$
and $l=-1$. The bottom four plots are cylindrical vector
modes: (c) the radial mode ($L=-1,S=1$, $\mathrm{TM}_{01}$); (d) the  azimuthal
mode ($L=1,S=-1$, $\mathrm{TE}_{01}$); and the hybrid modes, (e) even
($L=1,S=1$, $\mathrm{HE}_{e,21}$) and (f) odd ($L=-1,S=-1$, $\mathrm{HE}_{o,21}$).\label{fig:Transverse-intensity-profiles}}
\end{figure}

We begin by defining a general type of mode, those with cylindrically
symmetric intensity distributions:

\begin{equation}
\mathbf{M}^{[u]}(r,\phi)=R^{[m]}(r)\boldsymbol{\Phi}^{[n]}(\phi),\label{eq:modedef}
\end{equation}
where $r$ and $\phi$ are the standard radial and azimuthal cylindrical
coordinates, respectively and the position is denoted $\mathbf{r}=(r,\phi,z)$.
Throughout, quantities in bold font are vectors. Here, $\boldsymbol{\Phi}^{[n]}(\phi)$
gives the azimuthal dependence of the polarization (which is strictly
transverse) and phase for mode $n$. The radial dependence $R^{[m]}(r)$
is usually implicitly dependent on $n$ and will typically exhibit
a number of rings in the intensity profile that equal the radial mode
index $m$. We take $u$ to be the set of mode indices, i.e., $u=m,n$.
By defining $\mathbf{M}^{[u]}(r,\phi)$ in this way the intensity
distribution is explicitly cylindrically symmetric.

Inside a cylindrically symmetric waveguide with weak guiding (e.g.
with small refractive index contrast), azimuthally symmetric modes
are approximate paraxial solutions to the wave equation \citep{Zh2009}:

\begin{equation}
\nabla^{2}\mathbf{L}(\mathbf{r},t)+\frac{n(r)^{2}}{c^{2}}\frac{\partial^{2}\mathbf{L}(\mathbf{r},t)}{\partial t^{2}}\approx0.\label{eq:LinearSolution}
\end{equation}
Here, we have defined the normalized electric field solutions as 
\begin{equation}
\mathbf{L}^{[u]}(\mathbf{r},t)=\mathbf{M}^{[u]}(r,\phi)\exp(i\beta^{[u]}z)\exp(-i\omega t).
\end{equation}
Given a cylindrically symmetric index profile $n(r)$, one can find
an effective wavevector $\beta^{[u]}$ for each mode $\mathbf{M}^{[u]}$.
Both will depend on the angular frequency of the field $\omega$.
Fig. \ref{fig:Four-examples} shows a sample of common fiber structures
to which this work is applicable.

The $\mathbf{M}^{[u]}(r,\phi)$ modes obey the following orthonormality
relations \cite{SnyderLove:1983} for transversely polarized modes:
\begin{eqnarray}
\delta_{u,u'} & = & \int\mathbf{M}^{[u]*}(r,\phi)\cdot\mathbf{M}^{[u']}(r,\phi)r\mathrm{d}r\mathrm{d}\phi,\label{eq:ONCond}\\
\delta_{m,m'} & = & \int R^{[m]}(r)R^{[m']}(r)r\mathrm{d}r,\label{eq:ONCondR}\\
\delta_{n,n'} & = & \int\boldsymbol{\Phi}^{[n]*}(\phi)\cdot\boldsymbol{\Phi}^{[n']}(\phi)(r,\phi)\mathrm{d}\phi,\label{eq:ONCondphi}
\end{eqnarray}
where $\delta_{j,k}$ is the Kronecker delta. Here, the integral in Eq. (\ref{eq:ONCond})
is over the transverse plane and $u$ and $u'$ are composite indices
incorporating all the indices of the modes. Although Eq. (\ref{eq:ONCond}) is only strictly true for fields
of equal wavelength, we make the approximation that for small wavelength
differences the orthonormality is preserved. Eqs. (\ref{eq:ONCondR}) and (\ref{eq:ONCondphi}) follow from Eqs. (\ref{eq:ONCond}) and (\ref{eq:modedef}). This paper
focuses on the azimuthal mode function $\boldsymbol{\Phi}^{[n]}$,
which describes the spatial polarization variation of the modes. Since
this variation is independent of wavelength, the $\boldsymbol{\Phi}$
othonormality in Eq. (\ref{eq:ONCondphi}) will also be wavelength independent.
In the next three subsections, we will introduce three types of $\boldsymbol{\Phi}$
modes, each of which composes a complete mode-basis. Examples of the mode types are shown in Fig. \ref{fig:Transverse-intensity-profiles}. 

\subsection{Definite Spin and Orbital Angular Momentum Modes}

The azimuthal mode function $\boldsymbol{\Phi}^{[n]}(\phi)$ describes
both spin and orbital angular momentum. The first type of mode has
a definite value for both the spin and orbital angular momentum along
the system axis $z$ (e.g., the beam or fibre axis). The SAM of a photon
is given by its circular polarization, 
\begin{equation}
\boldsymbol{\sigma}^{[s]}=\frac{\left(\mathbf{x}+is\mathbf{y}\right)}{\sqrt{2}},\label{eq:spineqn}
\end{equation}
with right or left circular modes yielding a spin projection along
the system axis of $s\hbar$ ($s=\pm1$). The OAM results from an
azimuthal phase gradient of the field about the beam axis, $\exp(il\phi)$,
and has a value of $l\hbar$ ($l=0,\pm1,\pm2,\ldots)$ projected along
that same axis \citep{Woerdman1992}. With these functions, we can
define azimuthal modes $\boldsymbol{\Phi}^{[n]}(\phi)$ with definite
values of SAM and OAM, the $\mathbf{Y}$ modes: 
\begin{equation}
\mathrm{\mathbf{Y}}^{[l,s]}=e^{il\phi}\boldsymbol{\sigma}^{[s]}.\label{eq:ModeEx}
\end{equation}
In free-space, these are paraxial vector solutions to the wave equation.
In waveguides, these azimuthal modes are solutions only when $l\geq2$
\cite{SnyderLove:1983}.

\subsection{Cylindrical Vector Modes}

Our second type of $\boldsymbol{\Phi}$ mode is the cylindrical vector
mode. These are the general set of solutions in cylindrically-symmetric
weakly-guiding waveguides, valid for $l\geq1$ (unlike the $\mathbf{Y}$
modes). The CV azimuthal modes are:

\begin{equation}
\mathbf{CV}^{[L,S]}=\frac{1}{\sqrt{2}}\left(\mathrm{e}^{i\frac{\pi}{4}\left(S-1\right)}\mathrm{\mathbf{Y}}^{[L,S]}+\mathrm{e}^{-i\frac{\pi}{4}\left(S-1\right)}\mathrm{\mathbf{Y}}^{[-L,-S]}\right),\label{eq:CV_modes}
\end{equation}
where $S=\pm1$ and $L=\pm~\left|l\right|$ are mode indices. A unique
feature of these modes is that in contrast to the azimuthal modes
$\mathrm{\mathbf{Y}}^{[l,s]}$, they are real at all transverse points.
The CV modes cannot be factored into functions for OAM and SAM; hence,
they are non-separable and are no longer eigenstates of OAM or spin
(unlike the $\mathbf{Y}$ modes). Thus, for clarity, we will henceforth
refrain from referring to the spin of a particular CV mode. However,
the magnitude of $L$ does correspond to the magnitude of angular
momentum $l$ in each CV mode, $\left|L\right|=\left|l\right|$. 

The CV modes can be divided into groups of four that have effective
wavevector $\beta^{[u]}$ values that are close to each other \cite{SnyderLove:1983}.
Each mode in a particular mode group has an identical radial mode
function (i.e., index $m$) and identical orbital angular momentum
magnitude $\left|l\right|$:

\begin{flalign}
 & \mathbf{CV}^{[L=-\left|l\right|,S=1]}(\phi)=\frac{1}{\sqrt{2}}\left(\mathrm{\mathbf{Y}}^{[-\left|l\right|,1]}+\mathrm{\mathbf{Y}}^{[\left|l\right|,-1]}\right);\nonumber\\
 & \phantom{\mathbf{CV}}\phantom{\mathbf{CV}}\phantom{\mathbf{CV}}\phantom{\mathbf{CV}}\phantom{\mathbf{CV}}(\left|l\right|=1:\mathrm{TM}_{0m},\left|l\right|\geq2:\mathrm{EH}_{e,\left|l\right|-1,m})\label{eq:M1}\\
 & \mathbf{CV}^{[L=\left|l\right|,S=-1]}(\phi)=\frac{i}{\sqrt{2}}\left(\mathrm{\mathbf{Y}}^{[-\left|l\right|,1]}-\mathrm{\mathbf{Y}}^{[\left|l\right|,-1]}\right);\nonumber\\
 & \phantom{\mathbf{CV}}\phantom{\mathbf{CV}}\phantom{\mathbf{CV}}\phantom{\mathbf{CV}}\phantom{\mathbf{CV}}(\left|l\right|=1:\mathrm{TE}_{0m},\left|l\right|\geq2:\mathrm{EH}_{o,\left|l\right|-1,m})\label{eq:M2}\\
 & \mathbf{CV}^{[L=\left|l\right|,S=1]}(\phi)=\frac{1}{\sqrt{2}}\left(\mathrm{\mathbf{Y}}^{[-\left|l\right|,-1]}+\mathrm{\mathbf{Y}}^{[\left|l\right|,1]}\right);\nonumber\\
 & \phantom{\mathbf{CV}}\phantom{\mathbf{CV}}\phantom{\mathbf{CV}}\phantom{\mathbf{CV}}\phantom{\mathbf{CV}}(\left|l\right|\geq1:\mathrm{HE}_{e,\left|l\right|+1,m})\label{eq:M3}\\
 & \mathbf{CV}^{[L=-\left|l\right|,S=-1]}(\phi)=\frac{-i}{\sqrt{2}}\left(\mathrm{\mathbf{Y}}^{[-\left|l\right|,-1]}-\mathrm{\mathbf{Y}}^{[\left|l\right|,1]}\right);\nonumber\\
 & \phantom{\mathbf{CV}}\phantom{\mathbf{CV}}\phantom{\mathbf{CV}}\phantom{\mathbf{CV}}\phantom{\mathbf{CV}}(\left|l\right|\geq1:\mathrm{HE}_{o,\left|l\right|+1,m}).\label{eq:M4}
\end{flalign}
The four modes above have identical intensity distributions but different
patterns of spatially varying polarization. The set of modes for $|L|=1$
are shown in Fig. \ref{fig:Transverse-intensity-profiles}. This ladder
of modes are the mode solutions to cylindrically-symmetric weakly-guiding
waveguides. 

To understand the relationship of the CV modes to the Y modes as waveguide
solutions one must consider the degeneracy of the CV modes. Some of
the CV modes within each group of four will be degenerate. That is,
they will have equal $\beta^{[u]}$ within the weakly guiding approximation
in a waveguide. Notice that two of the four modes have OAM and SAM
aligned in each term, whereas the other two modes have anti-aligned
angular momenta. The aligned pair of modes in any weak-guiding cylindrically
symmetric fibers are degenerate, and likewise for the anti-aligned
pair. Any superposition of two degenerate modes is also a solution
and has the same $\beta^{[u]}$ as the original modes. Each $\mathrm{\mathbf{Y}}$
mode is a superposition of two degenerate CV modes and, consequently,
is a waveguide-mode solution. An exception is the case of $l=\pm1$
where the anti-aligned degeneracy is broken \citep{SnyderLove:1983,Gregg2015}.
i.e., $\mathbf{CV}^{[L=-1,S=1]}$ and $\mathbf{CV}^{[L=1,S=-1]}$
are not degenerate. In summary, whereas Y modes are solutions to cylindrically-symmetric
weakly-guiding waveguides for $l\geq2$, CV modes are solutions for
$l\geq1$.

\subsection{Total Angular Momentum Modes}

In addition to the $\mathrm{\mathbf{CV}}$ and $\mathrm{\mathbf{Y}}$
mode bases, it will be useful to define a $\boldsymbol{\Phi}$ mode
basis for the total angular momentum (TAM) projected along the beam
or fiber axis. In this basis, each mode has a definite TAM of $j=l+s$,
which will allow the total angular momentum conservation to be tracked
in the FWM processes. Since the FWM process occurs along the full
length of the medium, this is only useful to do if these $j$ eigenstates
are also eigenmodes of the fiber so that $j$ is conserved during
propagation. (Note, an eigenstate has a definite value of some observable
whereas an eigenmode remains unchanged upon propagation.) That is,
since both FWM and linear propagation occur concurrently it could
be the linear propagation rather than the FWM that causes $j$ to
change, obscuring the role of the FWM. 

For $\left|l\right|\geq2$, these definite TAM modes will be Y modes.
Since the Y modes have definite $s$ and $l$, the TAM $j$ will be
definite as well. However, as explained above, for the $\left|l\right|=1$
subspace, only two of the four Y modes are waveguide mode solutions,
$\mathrm{\mathbf{Y}}^{[1,1]}$ and $\mathrm{\mathbf{Y}}^{[-1,-1]}$.
For the other two Y modes $s$ and $l$ will not be preserved during
propagation. In their place, we use the anti-aligned pair of $\left|l\right|=1$
CV modes. Notice that both superposition terms in each of these CV
modes, Eqs. (\ref{eq:M1}) and (\ref{eq:M2}), have the same value for
TAM, $l+s=j=0$. Consequently, the anti-aligned CV modes are definite
TAM states with $j=0$. The TAM mode basis comprises these four modes,
which are listed in Table~\ref{Jmodes} and are represented by Z
in subsequent theory.

\begin{table}
\centering \captionsetup{justification=centering} %
\caption{Total Angular Momentum Modes. \label{Jmodes}}
\begin{tabular}{|c|c|}
\hline 
$l+s=j$  & Mode\tabularnewline
\hline 
$-1+1=0$  & $\mathbf{Z}^{[+0]}$=$\mathbf{CV}^{[L=-1,S=1]}$\tabularnewline
\hline 
$1-1=0$  & $\mathbf{Z}^{[-0]}=\mathbf{CV}^{[L=1,S=-1]}$\tabularnewline
\hline 
$1+1=2$  & $\mathrm{\mathbf{Z}}^{[2]}=\mathrm{\mathbf{Y}}^{[1,1]}$\tabularnewline
\hline 
$-1-1=-2$  & $\mathrm{\mathbf{Z}}^{[-2]}=\mathrm{\mathbf{Y}}^{[-1,-1]}$\tabularnewline
\hline 
\end{tabular}
\end{table}

\section{Four-Wave Mixing Theory for Spatial Light Modes}

We begin our FWM theory with the wave equation for a third-order nonlinear
process. Unlike most other published theory we retain the transverse
vector and spatial variation of the fields. As we shall discuss later,
we assume an isotropic material (such as silica) and that the nonlinearity
is frequency independent. This allows us to use a simplified form
for the third-order nonlinear polarization. In these systems (e.g.,
optical fiber), the coupled amplitude equations for the four fields
are dependent on the vector eigenmodes. In our theory, the effects
of self-phase modulation (SPM) and cross-phase modulation (XPM) arising
from the pump beams are also included. However, SPM and XPM arising
from the signal and idler are neglected on the basis that they are
far less intense than the pumps. Similarly, pump depletion is also
neglected in our treatment.

\subsection{Definition of Fields}

Four-wave mixing converts light from two pump $(p,p')$ fields to
two outgoing fields, typically called signal $(s)$ and idler $(i)$,
i.e., $p+p'\rightarrow s+i$. In the nondegenerate case, all four
fields can have distinct frequencies and spatial modes. We define
the total electric field vector in the system as the sum of the four
fields,

\begin{equation}
\mathbf{E}(\mathbf{r},t)=\underset{v=p,p',s,i}{\sum}\mathbf{E}_{v}(\mathbf{r},t).
\end{equation}
Each field is identified by the subscript $v$ and is assumed to be
monochromatic and exist in just one of the spatial modes discussed
in Section 2. Thus, this subscript will be taken to implicitly represent
the field identity $(p,p',i,$or $s)$, the field frequency $\omega_{v}$,
and the spatial mode $\mathbf{M}^{[u]}$. When considering an individual
field $\mathbf{E}_{v}$, we can factor out the slowly-varying field
amplitude $A_{v}(z)$:

\begin{equation}
\mathbf{E}_{v}(\mathbf{r},t)=A_{v}(z)\mathbf{L}^{[u]}(\mathbf{r},t)+\text{c.c.}.
\end{equation}
This definition will allow us to later isolate the slow change in
field amplitude $A_{v}(z)$ due to the nonlinear interaction.

\subsection{The Nonlinear Wave Equation}

The wave equation with a nonlinear source term $\mathbf{P}_{NL}$
is

\begin{equation}
\nabla^{2}\mathbf{E}(\mathbf{r},t)+\frac{n(r)^{2}}{c^{2}}\frac{\partial^{2}\mathbf{E}(\mathbf{r},t)}{\partial t^{2}}=-\mu_{0}\frac{\partial^{2}\mathbf{P}_{NL}(\mathbf{r},t)}{\partial t^{2}},\label{eq:wave1}
\end{equation}
where $\mu_{o}$ is the permeability of free space and $c$ is the
speed of light in vacuum. When considering all four fields pertinent
to FWM, the left-hand side of Eq. (\ref{eq:wave1}) is

\begin{multline}
\mathrm{LHS[Eq.}(\ref{eq:wave1})]\\
=\underset{v=p,p',s,i}{\sum}\left[\nabla^{2}\mathbf{E}_{v}(\mathbf{r},t)+\frac{n(r)^{2}}{c^{2}}\frac{\partial^{2}\mathbf{E}_{v}(\mathbf{r},t)}{\partial t^{2}}\right].\label{eq:wave1p2}
\end{multline}
Carrying out the spatial differentiation in the Laplacian, Eq. (\ref{eq:wave1})
becomes 
\begin{multline}
\mathrm{LHS[Eq.(\ref{eq:wave1})]}=\underset{v=p,p',s,i}{\sum}\Biggl[\left(\frac{\partial^{2}A_{v}(z)}{\partial z^{2}}+2i\beta_{v}\frac{\partial A_{v}(z)}{\partial z}\right)\mathbf{L}_{v}(\mathbf{r},t)\\
+A_{v}(z)\left(\nabla^{2}\mathbf{L}_{v}(\mathbf{r},t)+\frac{n(r)^{2}}{c^{2}}\frac{\partial^{2}\mathbf{L}_{v}(\mathbf{r},t)}{\partial t^{2}}\right)+\text{c.c.}\Biggl].
\end{multline}
In the right-hand side of the first line of this equation, we make
use of the fact that $A(z)$ is assumed to be slowly varying compared
to the fast oscillation associated with the effective wavevector $\beta^{[u]}$
along $z$. This is the slowly-varying amplitude approximation, ${\partial^{2}A(z)}/{\partial z^{2}}=0$.
The second line is just the source-free wave equation, Eq. (\ref{eq:LinearSolution}),
which is zero for the solution $\mathbf{L}_{v}(\mathbf{r},t)$. With
these simplifications, Eq. (\ref{eq:wave1}) becomes:

\begin{equation}
\underset{v=p,p',s,i}{\sum}\Biggl[2i\beta_{v}\frac{\partial A_{v}(z)}{\partial z}\mathbf{L}_{v}(\mathbf{r},t)+\text{c.c.}\Biggl]=-\mu_{0}\frac{\partial^{2}\mathbf{P}_{NL}(\mathbf{r},t)}{\partial t^{2}}.\label{eq:wave2}
\end{equation}

\subsection{Nonlinear Polarization and the Coupled Amplitude Equations}

We now determine the form of the nonlinear polarization $\mathbf{P}_{NL}(\mathbf{r},t)$.
We omit contributions from other third-order processes such as third
harmonic generation, etc. By considering only the nonlinear polarization
$\mathbf{P}_{v}(\mathbf{r})$ created at the pump, signal, and idler
frequencies, the nonlinear polarization can be expressed as the sum

\begin{equation}
\mathbf{P}_{NL}(\mathbf{r},t)=\underset{v=p,p',s,i}{\sum}\mathbf{P}_{v}(\mathbf{r})\exp(-i\omega_{v}t)+\text{c.c.}.
\end{equation}
With this definition, the right-hand side of the nonlinear wave equation,
Eq. (\ref{eq:wave2}), becomes

\begin{equation}
\mathrm{RHS[Eq.(\ref{eq:wave2})]}=\mu_{0}\underset{v=p,p',s,i}{\sum}\omega_{v}^{2}\mathbf{P}_{v}(\mathbf{r})\exp(-i\omega_{v}t)+\text{c.c.}.
\end{equation}
In materials that are both isotropic and satisfy the Kleinman symmetry condition
(i.e., the frequency dependence of $\chi^{(3)}$ can be neglected
\citep{Boyd:2008:NOT:1817101}), the third order nonlinear susceptibility
tensor can be expressed in terms of a single scalar component (i.e.
$\chi_{xxxx}^{(3)}$). As we show in Appendix A, the nonlinear polarization
at the signal frequency is then given by

\begin{multline}
\mathbf{P}_{s}(\mathbf{r})=2\epsilon_{0}\chi_{xxxx}^{(3)}\biggl[\\
\left(\mathbf{E}_{p}^{*}\cdot\mathbf{E}_{p}\right)\mathbf{E}_{s}+\left(\mathbf{E}_{p}^{*}\cdot\mathbf{E}_{s}\right)\mathbf{E}_{p}+\left(\mathbf{E}_{p}\cdot\mathbf{E}_{s}\right)\mathbf{E}_{p}^{*}\\
+\left(\mathbf{E}_{p'}^{*}\cdot\mathbf{E}_{p'}\right)\mathbf{E}_{s}+\left(\mathbf{E}_{p'}^{*}\cdot\mathbf{E}_{s}\right)\mathbf{E}_{p'}+\left(\mathbf{E}_{p'}\cdot\mathbf{E}_{s}\right)\mathbf{E}_{p'}^{*}\\
+\left(\mathbf{E}_{i}^{*}\cdot\mathbf{E}_{p}\right)\mathbf{E}_{p'}+\left(\mathbf{E}_{i}^{*}\cdot\mathbf{E}_{p'}\right)\mathbf{E}_{p}+\left(\mathbf{E}_{p}\cdot\mathbf{E}_{p'}\right)\mathbf{E}_{i}^{*}\biggr],\label{eq:nlp1}
\end{multline}
where we have omitted the spatial-temporal coordinates for clarity.
A similar equation can be found for the nonlinear polarization at
the idler frequency $\mathbf{P}_{i}$ by exchanging $s$ and $i$
throughout Eq. (\ref{eq:nlp1}). The first six terms in Eq. (\ref{eq:nlp1})
represent XPM from the pump beams and the last three terms are the
contribution to FWM.

Now that we have a succinct form for the nonlinear polarization, we
can evaluate its action on the signal and idler fields. The nonlinear
wave equation (Eq. (\ref{eq:wave2})), for just the signal field is
\begin{gather}
2i\beta_{s}\frac{\partial A_{s}(z)}{\partial z}\mathbf{M}_{s}(r,\phi)\exp(i\beta_{s}z)=\frac{\omega_{s}^{2}\mathbf{P}_{s}(\mathbf{r})}{\epsilon_{0}c^{2}}.\label{eq:nls}
\end{gather}
Substituting in the nonlinear polarization of the signal frequency
$\mathbf{P}_{s}$ from Eq. (\ref{eq:nlp1}) into Eq. (\ref{eq:nls})
above, we obtain 
\begin{multline}
\frac{\partial A_{s}(z)}{\partial z}\mathbf{M}_{s}=\frac{-i\omega_{s}^{2}\chi_{xxxx}^{(3)}}{2\beta_{s}c^{2}}\Biggl\{\\
A_{p}^{*}A_{p}A_{s}\biggl[\boldsymbol{\alpha}_{pps}+\boldsymbol{\alpha}_{psp}+\bar{\boldsymbol{\alpha}}_{psp}\biggr]\\
+A_{p'}^{*}A_{p'}A_{s}\biggl[\boldsymbol{\alpha}_{p'p's}+\boldsymbol{\alpha}_{p'sp'}+\bar{\boldsymbol{\alpha}}_{p'sp'}\biggr]\\
+A_{p}A_{p'}A_{i}^{*}\biggl[\boldsymbol{\alpha}_{ipp'}+\boldsymbol{\alpha}_{ip'p}+\bar{\boldsymbol{\alpha}}_{pp'i}\biggr]\exp(i\Delta kz)\Biggr\},
\end{multline}
where $\Delta k=\beta_{p}+\beta_{p'}-\beta_{s}-\beta_{i}$ is the
phasematching term and where we introduce the quantities $\boldsymbol{\alpha}_{ijk}=\left(\mathbf{M}_{i}^{*}\cdot\mathbf{M}_{j}\right)\mathbf{M}_{k}$
and $\boldsymbol{\bar{\alpha}}_{ijk}=\left(\mathbf{M}_{i}\cdot\mathbf{M}_{j}\right)\mathbf{M}_{k}^{*}$.
As before, we have suppressed the dependence of $A$ and $\mathbf{M}$
on the spatial coordinates.

We now isolate the behaviour of the scalar amplitudes $A$ by taking
the dot product with $\mathbf{M}_{s}^{*}$ on both sides and integrating
over the transverse plane. With this, the orthonormality condition
in Eq.~(\ref{eq:ONCond}) yields the coupled amplitude equation for
the signal (or idler, by exchanging $s$ and $i$ throughout) as

\begin{multline}
\frac{\partial A_{s}(z)}{\partial z}=\kappa_{s}\Bigl\{\Bigl|A_{p}\Bigr|^{2}A_{s}\left[\textsf{O}_{p^{*}ps^{*}s}+\textsf{O}_{p^{*}ss^{*}p}+\textsf{O}_{pss^{*}p^{*}}\right]\\
+\Bigl|A_{p'}\Bigr|^{2}A_{s}\left[\textsf{O}_{p'^{*}p's^{*}s}+\textsf{O}_{p'^{*}ss^{*}p'}+\textsf{O}_{p'ss^{*}p'^{*}}\right]\\
+A_{p}A_{p'}A_{i}^{*}\left[\textsf{O}_{i^{*}ps^{*}p'}+\textsf{O}_{i^{*}p's^{*}p}+\textsf{O}_{pp's^{*}i^{*}}\right]\exp(i\Delta kz)\Bigr\}.\label{eq:coupled}
\end{multline}
Here, the field coupling constant $\kappa_{s}$ is given by $\kappa_{s}=\frac{-i\omega_{s}^{2}\chi_{xxxx}^{(3)}}{2\beta_{s}c^{2}}$,
and $\textsf{O}_{a^{(*)}b^{(*)}c^{(*)}d^{(*)}}=\int\left(\mathbf{M}_{a}^{(*)}\cdot\mathbf{M}_{b}^{(*)}\right)\left(\mathbf{M}_{c}^{(*)}\cdot\mathbf{M}_{d}^{(*)}\right)r\mathrm{d}r\mathrm{d}\phi$
is the mode overlap integral. In $\textsf{O}_{a^{(*)}b^{(*)}c^{(*)}d^{(*)}}$,
the subscripts $a,b,c,$ and $d$ represent the fields ($p$, $p'$,
$s$, or $i$) and an optional conjugate on a particular subscript
applies to the corresponding mode in the integral. For example, $\textsf{O}_{i^{*}ps^{*}p'}=\int\left(\mathbf{M}_{i}^{*}\cdot\mathbf{M}_{p}\right)\left(\mathbf{M}_{s}^{*}\cdot\mathbf{M}_{p'}\right)r\mathrm{d}r\mathrm{d}\phi$.
So far, we have not used the cylindrical symmetry of the modes. This
coupled wave equation is applicable in bulk media and waveguides since
it only assumes that $\mathbf{M}_{v}$ are paraxial modes that are
approximate solutions to the wave equation. (In Appendix A, we give
the analogous coupled amplitude equation for the pump field.)

We now consider the conditions necessary for efficient four-wave mixing.
The first two lines of the RHS of Eq. (\ref{eq:coupled}) describe
XPM in the signal field induced by the pumps. The last line describes
four-wave mixing. Four-wave mixing requires perfect energy conservation
and is most efficient when photon momentum is also conserved. These
two conditions constitute perfect phasematching, $\Delta\omega=\omega_{p}+\omega_{p'}-\omega_{i}-\omega_{s}=0$
and $\Delta k=0$, which can occur in many different FWM processes
since $\beta_{v}$ varies with mode and frequency \citep{Boyd:2008:NOT:1817101}.
Since the energy and momentum conservation for FWM will depend on
the specific medium and waveguide of interest, we will not further
consider the details of phasematching in this paper.

The selection rules will arise from a third condition for FWM. Namely,
in the final line of the coupled wave equation, Eq. (\ref{eq:coupled}),
the total process amplitude, 
\begin{equation}
\textsf{U}\equiv\textsf{O}_{i^{*}ps^{*}p'}+\textsf{O}_{i^{*}p's^{*}p}+\textsf{O}_{pp's^{*}i^{*}},\label{eq:FWMterms}
\end{equation}
must be non-zero.

\section{Four-wave Mixing Selection Rules}

So far, the coupled wave equations have been derived for any transverse
modes satisfying the linear wave equation. We now restrict ourselves
to systems with cylindrical symmetry, such as optical fibers. We introduce
cylindrically symmetric modes into the FWM integrals listed in Eq.
(\ref{eq:FWMterms}) in order to find a set of selection rules based
on the transitions permitted between the modes. To start, we use the
mode definition in Eq. (\ref{eq:modedef}) to separate the overlap
integrals comprising the process amplitude $\textsf{U}$ into a product
of radial and azimuthal parts such that

\begin{equation}
\textsf{O}_{a^{(*)}b^{(*)}c^{(*)}d^{(*)}}=F\int\left(\boldsymbol{\Phi}_{a}^{(*)}\cdot\boldsymbol{\Phi}_{b}^{(*)}\right)\left(\boldsymbol{\Phi}_{c}^{(*)}\cdot\boldsymbol{\Phi}_{d}^{(*)}\right)\text{d}\phi,\label{eq:modeoverlap}
\end{equation}
where $F\equiv\int R_{i}^{*}R_{p}R_{s}^{*}R_{p'}r\mathrm{d}r$ is
the integral over the radial dependence. Since the radial mode function
$R_{v}=R^{[m(v)]}(r)$ depends on the specific radial index profile
$n(r)$ of a chosen waveguide, we will focus on the selection rules
set by the azimuthal modes. These selection rules will then be general
to any medium with cylindrical symmetry. Specifically we assume that
$F$ is non-zero, as would be true if all four fields were in the
same radial mode (i.e., the same value of $m$). Henceforth, a ``mode''
will refer to only its azimuthal component $\boldsymbol{\Phi}_{v}$($\phi)$.
(Note that for the CV mode basis only, the potential complex conjugates
of $\boldsymbol{\Phi}_{v}$ in the integral Eq. (\ref{eq:modeoverlap})
can be dismissed since at all transverse positions the modes are real.)

The total FWM intensity is proportional to the sum $\textsf{U}$,
Eq.~(\ref{eq:FWMterms}), of three overlap integrals, each of which
is defined by Eq.~(\ref{eq:modeoverlap}). By considering the mode
combinations for the four fields which lead to a non-zero $\textsf{U}$,
we can find the corresponding sets of selection rules for a given
set of mode indices. We will express allowed FWM processes in the
notation, $\mathbf{M}_{p}+\mathbf{M}_{p'}\rightarrow\mathbf{M}_{i}+\mathbf{M}_{s}$.
All allowed processes can also occur in reverse, i.e., with input
and output modes interchanged. If a process occurs for a particular
set of modes in fields $p$ and $p'$ then it will also occur with
the modes interchanged between $p$ and $p'$ and likewise for output
fields $s$ and $i$. 

Of particular interest will be the scenario where all the involved
modes are taken from the family of four modes (within the Y, CV, and
Z mode types) that have equal $|l|$. With this limited set of four,
there are a finite number of potential processes, which we will determine.

\subsection{Summary of the Y Mode Selection Rules}
\begin{table*}[bt]
\centering \captionsetup{justification=centering} %
\caption{Selection Rules for CV Modes.\label{CVselectrule1}}
\begin{tabular}{|c|c|c|c|}
\hline 
Rule  & $L$  & $S$  & Process Amplitude, $\textsf{U}$ \tabularnewline
\hline 
1  & $L_{p}+L_{p'}=L_{s}+L_{i}$  & $S_{p}+S_{p'}=S_{s}+S_{i}$  & $2\pi F$ \tabularnewline
\hline 
2  & $L_{p}+L_{p'}=-\left(L_{s}+L_{i}\right)$  & $S_{p}+S_{p'}=-\left(S_{s}+S_{i}\right)$  & $-2\pi F$ \tabularnewline
\hline 
3  & $L_{p}-L_{p'}=L_{s}-L_{i}$  & $S_{p}-S_{p'}=S_{s}-S_{i}$  & $2\pi F$ \tabularnewline
\hline 
4  & $L_{p}-L_{p'}=-\left(L_{s}-L_{i}\right)$  & $S_{p}-S_{p'}=-\left(S_{s}-S_{i}\right)$  & $2\pi F$ \tabularnewline
\hline 
\end{tabular}
\end{table*}
For the sake of completeness and for use later, we derive the selection
rules for fields in the Y angular momentum modes. Our results agree
with similar FWM studies involving linearly polarized modes~\cite{Garay2016}.
We insert combinations of Y modes into the three $\textsf{O}$ overlap
integrals inside $\textsf{U}$ (Eq. (\ref{eq:FWMterms})). Each overlap
integral factors into OAM and SAM components. For example, 
\begin{eqnarray}
\textsf{O}_{i^{*}ps^{*}p'} & = & F\left(\boldsymbol{\sigma}^{[-s_{i}]}\cdot\boldsymbol{\sigma}^{[s_{p}]}\right)\left(\boldsymbol{\sigma}^{[-s_{s}]}\cdot\boldsymbol{\sigma}^{[s_{p'}]}\right)\nonumber \\
 & \times & \int e^{i\left(-l_{i}+l_{p}-l_{s}+l_{p'}\right)\phi}\text{d}\phi\nonumber \\
 & = & F\delta_{s_{i},s_{p}}\delta_{s_{s},s_{p'}}\delta_{l_{p}+l_{p'},l_{s}+l_{i}}.\label{eq:Oexample}
\end{eqnarray}
The other two $\textsf{O}$ integrals give similar results. Given
that $s$ is limited to the values $\pm1$, the Kronecker deltas for
the $s$ values in the three integrals can be combined to arrive at
the selection rules: 
\begin{equation}
\begin{aligned}s_{p}+s_{p'}=s_{i}+s_{s},\\
l_{p}+l_{p'}=l_{i}+l_{s},
\end{aligned}
\label{eq:lsrules}
\end{equation}
where $l_{v}$ and $s_{v}$ are the mode indices of fields $v=p,p',s,i$. When these rules are satisfied, the process
will occur with amplitude $\textsf{U}=4\pi F$. We give a more detailed
derivation of this in Appendix~B.

These two rules show that SAM and OAM are independently conserved.
That is, there is no coupling between the SAM and OAM degrees of freedom.
Consequently, the only way for the total angular momentum along the
fiber or beam axis, $j=l+s,$ to be conserved is for SAM to be conserved
and OAM to be conserved: 
\begin{equation}
\begin{aligned}\Delta s & =s_{p}+s_{p'}-s_{s}-s_{i}=0,\\
\Delta l & =l_{p}+l_{p'}-l_{s}-l_{i}=0,\\
\Delta j & =j_{p}+j_{p'}-j_{s}-j_{i}=0.
\end{aligned}
\label{eq:jrules}
\end{equation}
In other words, in the weakly guiding or paraxial approximation, total
angular momentum cannot be conserved by converting SAM into OAM. This
is a result of the fact that in these approximations, the fields are
transverse everywhere.

\subsection{Summary of the CV Mode Selection Rules}

We now consider the situation in which each of the four fields is
in a CV mode. Each mode can have any value of $\left|L\right|$. We
label the mode indices for field $v=p,p',s,i$ as $L_{v}$ and $S_{v}.$
Table \ref{CVselectrule1} below gives the four selection rules from
our derivation, which is detailed in Appendix~B. More than one rule
can hold true simultaneously and their amplitudes should be added
to find the total process amplitude $\textsf{U}$. Since $S$
is limited to $\pm1$, the selection rules disallow processes where
there are an odd number of like $S$ values amongst $i,s,p,$ and
$p'$.

\begin{figure}
\includegraphics[width=0.9\columnwidth]{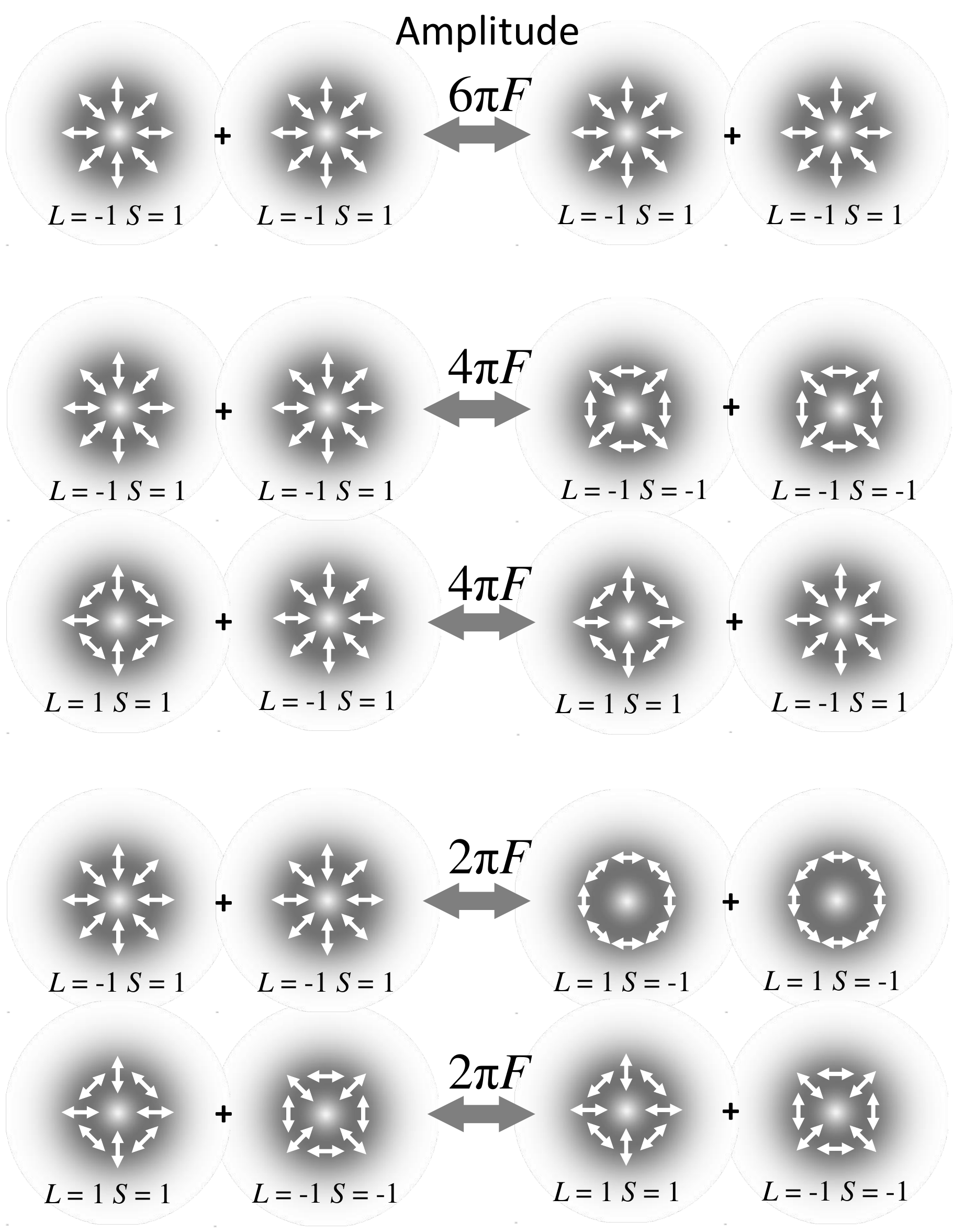}

\caption{Examples of allowed four-wave mixing processes between cylindrical vector modes.\label{fig:ConditionExamples}}
\end{figure}

We now consider the limited situation in which all the modes have
the same value of $|L|$. We compile all 100 distinct possible processes in
a table in Appendix B. Now, similar to the situation for $S$, since
$L$ is limited to $\pm|L|$, the selection rules disallow processes
where there are an odd number of like $L$ values amongst $i,s,p,$
and $p'$. This is confirmed by the table, which shows that processes
where any given mode (e.g., TE) does not appear in pairs is ruled
out. For example, $\mathrm{TM}+\mathrm{TE}\rightarrow\mathrm{TM}+\mathrm{TE}$
or $\mathrm{HE}_{e}+\mathrm{HE}_{e}\rightarrow\mathrm{HE}_{e}+\mathrm{HE}_{e}$
or $\mathrm{TM}+\mathrm{TM}\rightarrow\mathrm{HE}_{o}+\mathrm{HE}_{o}$
are allowed but $\mathrm{TM}+\mathrm{TE}\rightarrow\mathrm{TM}+\mathrm{TM}$
is not. This single selection rule is similar to the selection rules
for linear polarized fields in an isotropic medium \cite{Agrawal2007}
that involve only two orthogonal polarizations. Examples of possible
CV mode conversions for total amplitudes ranging from $2\pi F$ to
$6\pi F$ are shown in Fig. \ref{fig:ConditionExamples}.

When considering higher order CV modes beyond the $|L|=1$ manifold,
one can find non-trivial allowed FWM processes such as $L_{p}=2$,
$L_{p'}=2$, $L_{s}=3$, and $L_{i}=1$ among others. However, since
radial functions $R^{[m]}(r)$ typically implicitly depend on $|l|$,
they will have different shapes for each mode. In turn, this will
decrease the radial overlap integral $F$ and, thus, these processes
will occur with a decreased efficiency.

\begin{table*}[ht]
\begin{centering}
\captionsetup{justification=centering} %
\caption{Selection Rules for Definite Total Angular Momentum Modes.\label{TAMselectrule}}
\begin{tabular}{|c|c|c|}
\hline 
Process Type  & Allowed Processes  & Amplitude, $\textsf{U}^{(\mathrm{TAM})}$\tabularnewline
\hline 
$\mathrm{\mathbf{Z}}^{[w0]}+\mathrm{\mathbf{Z}}^{[w0]}\rightarrow\mathrm{\mathbf{Y}}^{[q,q]}+\mathrm{\mathbf{Y}}^{[-q,-q]}$  & $w=\pm,q=\pm1$  & $4\pi F$\tabularnewline
\hline 
$\mathrm{\mathbf{Y}}^{[q,q]}+\mathrm{\mathbf{Y}}^{[-q,-q]}\rightarrow\mathrm{\mathbf{Z}}^{[w0]}+\mathrm{\mathbf{Z}}^{[w0]}$  & $w=\pm,q=\pm1$  & $4\pi F$\tabularnewline
\hline 
$\mathrm{\mathbf{Z}}^{[w0]}+\mathrm{\mathbf{Y}}^{[q,q]}\rightarrow\mathrm{\mathbf{Z}}^{[w0]}+\mathrm{\mathbf{Y}}^{[q,q]}$  & $w=\pm,q=\pm1$  & $4\pi F$\tabularnewline
\hline 
\end{tabular}
\par\end{centering}

\end{table*}
\subsection{Selection Rules for the TAM mode basis}

We now consider the situation in which each of the four fields is
in a TAM mode. The TAM mode set consists of the two $j=0$ modes,
$\mathbf{Z}^{[-0]}=\mathbf{CV}^{[-1,1]}$ and $\mathbf{Z}^{[+0]}=\mathbf{CV}^{[1,-1]}$
and the $j=\pm2$ modes $\mathrm{\mathbf{Y}}^{[1,1]}$ and $\mathrm{\mathbf{Y}}^{[-1,-1]}$.
A derivation of the relevant selection rules is given in Appendix
B. Here, we summarize them and consider if conservation of TAM is
enforced by the rules. In particular, we ask whether the sum of the
total angular momentum of the two input photons is conserved in four-wave
mixing, $j_{i}+j_{s}=j_{p}+j_{p'}$. 

Of the 100 potential processes, the ones involving solely Y modes
or solely Z modes are already treated by the selection rules in the
last two sections. We have already shown that the selection rules
for the Y modes conserve $j$ since they separately conserve OAM and
SAM. However, the Z modes are not eigenstates of OAM or SAM and, thus,
one cannot ascribe $s$ and $l$ values before and after the process.
It follows that it becomes impossible to evaluate their conservation.
Instead, we directly consider the TAM, $j$. A process containing
only Z modes trivially conserves the two photons' total angular momentum
given that each mode carries none. Thus, we focus on the non-trivial
processes that interconvert $j=0$ and $j=\pm2$ modes, which we summarize
in Table \ref{TAMselectrule}.

Considering the conservation of angular momentum, the first allowed
process reads $j_{i}+j_{s}-j_{p}-j_{p'}=\mp2\pm2-0-0=0$. The second
process, the reverse of the first, conserves TAM in a similar manner.
The last process reads $j_{i}+j_{s}-j_{p}-j_{p'}=0\pm2-0\mp2=0$.
Thus, all the TAM interconversion processes conserve total angular
momentum of the two photons.

One might ask whether conservation of total angular momentum could
be used as the sole selection rule. The answer is no. There are many
processes that would conserve TAM that are not permitted, such as
$\mathrm{\mathbf{Z}}^{[+0]}+\mathrm{\mathrm{\mathbf{Y}}^{[\pm1,\pm1]}}\rightarrow\mathrm{\mathbf{Z}}^{[-0]}+\mathrm{\mathrm{\mathbf{Y}}^{[\pm1,\pm1]}}$. Allowed processes can
occur with different amplitudes of $\textsf{U}=2\pi F,4\pi F,\textrm{ and }6\pi F$.

\subsection{Prospects for the generation of mode-entangled photon pairs}

Previous work has generated photon pairs that are entangled in their
OAM modes \cite{Fickler2012}. In those cases, entanglement occurred
naturally through the conservation of OAM. More generally, to generate
entanglement through a spontaneous nonlinear process such as downconversion
or four-wave mixing, one requires two simultaneous processes that
each produce a different pair of signal and idler modes. In order
to generate entanglement, the modes containing the signal and idler
photons must be indistinguishable other than in the degree of freedom
that is entangled \cite{Kwiat1995}. For FWM in a fiber, this entails
phasematching the two processes with identical signal wavelengths
and identical idler wavelengths. Since each mode has a different effective
wavevector $\beta$ with different dependence on wavelength, $\Delta k$
will typically vary differently with wavelength for the two processes.
Consequently, such phasematching is nontrivial. Fortunately, control
of the effective index $\beta$ can be achieved by careful design
of fiber index profile and choice of material \cite{Cohen2009}. 

In order for these two processes together to create mode entanglement,
they must produce different pairs of modes, $\mathrm{\mathbf{M}}_{i}$
and $\mathrm{\mathbf{M}}_{s}$, from one another. An example of two
appropriate processes is $\mathrm{\mathbf{Y}}_{p}^{[1,1]}+\mathrm{\mathbf{Y}}_{p'}^{[-1,-1]}\rightarrow\mathrm{\mathbf{Z}}_{i}^{[+0]}+\mathrm{\mathbf{Z}}_{s}^{[+0]}$
and $\mathrm{\mathbf{Y}}_{p}^{[1,1]}+\mathrm{\mathbf{Y}}_{p'}^{[-1,-1]}\rightarrow\mathrm{\mathbf{Z}}_{i}^{[-0]}+\mathrm{\mathbf{Z}}_{s}^{[-0]}$.
The entangled state of the two spontaneously generated photons would
be $\mathrm{\mathbf{Z}}_{i}^{[+0]}\mathrm{\mathbf{Z}}_{s}^{[+0]}+\mathrm{\mathbf{Z}}_{i}^{[-0]}\mathrm{\mathbf{Z}}_{s}^{[-0]}$.
The availability of many allowed four-wave mixing processes for cylindrical
vector modes means that there are many possibilities for the generation
of photons entangled in their spatial-polarization modes.

\section{Conclusion}

In conclusion, we present a theoretical investigation into the nonlinear
optics of structured light. Specifically, we developed a theory describing
four-wave mixing of cylindrical vector modes in optical fiber or free
space and the mode selection rules that follow. For comparison, we
also review analogous theory for modes with definite spin and orbital
angular momentum. Given the cylindrical symmetry of a fiber and free
space, one might expect that the only quantity that must be strictly
conserved in four-wave mixing is the total angular momentum along
the system axis. Indeed, in some processes the orbital and spin angular
momentum are independently conserved and, thus, so too will be the
total angular momentum. However, for orbital angular momentum $|l|=1,$
the fundamental eigenmodes of an optical fiber are not eigenstates
of either spin or orbital angular momentum and, consequently, these
properties will change during propagation. Nonetheless, we showed
that for modes that are total angular momentum eigenstates, the sum
of two photons' total angular momentum is conserved and emerges in
the output photons. Future work will investigate the link between
four-wave mixing selection rules and angular momentum conservation
laws outside of the paraxial and weakly-guiding regime. 

Since cylindrical vector modes are the eigenmodes of weakly guiding
fibers and of free-space propagation, the selection rules presented
here are pertinent to a range of applications. These include telecommunications,
where structured modes can increase bandwidths \citep{Bozinovic20131545,Liu2018}.
Additionally our results could find use in all-optical switching by
utilizing the optical nonlinearity of structured spatial modes as
a way to route such beams. In quantum optics, cylindrical vector modes
are currently being investigated for use in quantum cryptography \citep{Ndagano2017}.
Using our results, the CV mode photons could be generated directly
by FWM inside optical fibers. We also show how to generate photon
pairs entangled in their CV modes. Beyond fiber optics, our results
have implications in free-space four-wave mixing, and could shed light
on the debate over the roles of spin and orbital angular momentum
in photons \citep{Allen2003,Andrews2013}.

This work was supported by the Canada Research Chairs (CRC) Program,
the Canada First Research Excellence Fund (CFREF), and the Natural
Sciences and Engineering Research Council (NSERC), and the NRC-uOttawa
Joint Centre for Extreme Photonics (JCEP).

\section{Appendix}

\subsection{Derivation of the functional form of the Nonlinear Polarization}

In this appendix, we will derive the form of the third-order nonlinear
polarization $\mathbf{P}$ for an isotropic material. In general,
each component $P_{i}$ ($i=x,y,z$) of the $\chi^{(3)}$ nonlinear
polarization for a given process is \citep{Boyd:2008:NOT:1817101}

\begin{equation}
P_{i}=\epsilon_{0}D\sum_{jkl=x,y,z}\chi_{ijkl}^{(3)}E_{j}(\omega_{o})E_{k}(\omega_{n})E_{l}(\omega_{m}),\label{eq:NLPBoyd}
\end{equation}
where the degeneracy factor $D$ is equal to the number of distinct
permutations of the fields. For the moment, all subscripts refer to
Cartesian coordinates $x,y,z$ unlike in the main body of the paper,
where subscripts $p,p',s,$and $i$ identify the field.

The nonzero components of the $\chi^{(3)}$ tensor are

\begin{multline}
\chi_{xxyy}^{(3)}=\chi_{xxzz}^{(3)}=\chi_{yyxx}^{(3)}=\chi_{yyzz}^{(3)}=\chi_{zzxx}^{(3)}=\chi_{zzyy}^{(3)}\\
=\chi_{xyxy}^{(3)}=\chi_{xzxz}^{(3)}=\chi_{yzyz}^{(3)}=\chi_{yxyx}^{(3)}=\chi_{zxzx}^{(3)}=\chi_{zyzy}^{(3)}\\
=\chi_{xyyx}^{(3)}=\chi_{xzzx}^{(3)}=\chi_{yxxy}^{(3)}=\chi_{yzzy}^{(3)}=\chi_{zxxz}^{(3)}=\chi_{zyyz}^{(3)}\\
=\frac{1}{3}\chi_{xxxx}^{(3)}=\frac{1}{3}\chi_{yyyy}^{(3)}=\frac{1}{3}\chi_{zzzz}^{(3)}.\label{eq:Chi3}
\end{multline}

Since most of the $\chi^{(3)}$ tensor components are zero, for each
nonlinear polarization component $P_{i}$, only seven terms of the
27-term sum are nonzero. The last line of Eq. (\ref{eq:Chi3}) shows
that the $jkl=iii$ term is multiplied by a factor of three. Instead,
we divide the $iii$ term into three separate terms in order to make
evident an upcoming vector product. All together then, there are nine
nonzero terms in the nonlinear polarization: 
\begin{eqnarray}
 & P_{i}=\frac{\epsilon_{0}D\chi_{xxxx}^{(3)}}{3}\biggl[\nonumber \\
 & \biggl(E_{i}(\omega_{o})E_{i}(\omega_{n})+E_{j}(\omega_{o})E_{j}(\omega_{n})+E_{k}(\omega_{o})E_{k}(\omega_{n})\biggr)E_{i}(\omega_{m})\nonumber \\
+ & \biggl(E_{i}(\omega_{n})E_{i}(\omega_{m})+E_{j}(\omega_{n})E_{j}(\omega_{m})+E_{k}(\omega_{n})E_{k}(\omega_{m})\biggr)E_{i}(\omega_{o})\nonumber \\
+ & \biggl(E_{i}(\omega_{o})E_{i}(\omega_{m})+E_{j}(\omega_{o})E_{j}(\omega_{m})+E_{k}(\omega_{o})E_{k}(\omega_{m})\biggr)E_{i}(\omega_{n})\biggr]\nonumber \\
 & =\frac{\epsilon_{0}D\chi_{xxxx}}{3}\biggl[\nonumber \\
 & \Bigl(\mathbf{E}(\omega_{o})\cdot\mathbf{E}(\omega_{n})\Bigr)E_{i}(\omega_{m})\nonumber \\
 & \Bigl(\mathbf{E}(\omega_{m})\cdot\mathbf{E}(\omega_{n})\Bigr)E_{i}(\omega_{o})\nonumber \\
 & \Bigl(\mathbf{E}(\omega_{m})\cdot\mathbf{E}(\omega_{o})\Bigr)E_{i}(\omega_{n})\biggr],
\end{eqnarray}
where $i$, $j$ and $k$ are any permutation of $\left\{ x,y,z\right\} $.
We have also assumed that the frequency dependence of the susceptibility
can be neglected (the Kleinman symmetry condition). Writing the three
polarization Cartesian components as a vector we have

\begin{multline}
\mathbf{P}=\frac{\epsilon_{0}D\chi_{xxxx}^{(3)}}{3}\biggl[\Bigl(\mathbf{E}(\omega_{o})\cdot\mathbf{E}(\omega_{n})\Bigr)\mathbf{E}(\omega_{m})\\
+\Bigl(\mathbf{E}(\omega_{m})\cdot\mathbf{E}(\omega_{n})\Bigr)\mathbf{E}(\omega_{o})+\Bigl(\mathbf{E}(\omega_{m})\cdot\mathbf{E}(\omega_{o})\Bigr)\mathbf{E}(\omega_{n})\biggr].
\end{multline}
In the main text, the nonlinear polarization in Eq. (\ref{eq:nlp1})
follows from applying this to FWM and XPM (conjugating fields with
negative frequency) and evaluating the degeneracy factor: $D=6$ for
three distinct fields and $D=3$ for two distinct fields.

For completeness, the nonlinear polarization for the pump $p$ (or
$p'$, by exchanging $p$ and $p'$ throughout) is: 
\begin{multline}
\mathbf{P}_{p}(\mathbf{r})=\epsilon_{0}\chi_{xxxx}^{(3)}\biggl[\left(\mathbf{E}_{p}\cdot\mathbf{E}_{p}\right)\mathbf{E}_{p}^{*}+2\left(\mathbf{E}_{p}^{*}\cdot\mathbf{E}_{p}\right)\mathbf{E}_{p}\\
+2\left(\mathbf{E}_{p'}^{*}\cdot\mathbf{E}_{p'}\right)\mathbf{E}_{p}+2\left(\mathbf{E}_{p'}\cdot\mathbf{E}_{p}\right)\mathbf{E}_{p'}^{*}+2\left(\mathbf{E}_{p'}^{*}\cdot\mathbf{E}_{p}\right)\mathbf{E}_{p'}\biggr].\label{eq:nlp2}
\end{multline}
Here, the subscripts on the electric fields and polarization indicate
the field identity, $p$ or $p'$. By the same procedure used in the
main body of the paper, the coupled equations for the pump amplitude
$p$ (and $p'$, by exchanging $p$ and $p'$ throughout) are found
to be 
\begin{multline}
\frac{\partial A_{p}(z)}{\partial z}=\frac{1}{2}\kappa_{p}\Bigl\{\Bigl|A_{p}\Bigr|^{2}A_{p}\left[\textsf{O}_{ppp^{*}p^{*}}+2\textsf{O}_{p^{*}pp^{*}p}\right]\\
+2\Bigl|A_{p'}\Bigr|^{2}A_{p}\left[\textsf{O}_{p'^{*}p'p^{*}p}+\textsf{O}_{p'pp^{*}p'^{*}}+\textsf{O}_{p'^{*}pp^{*}p'}\right]\Bigr\}.\label{eq:coupledPump}
\end{multline}

\subsection{Calculation of Four-wave Mixing Process Amplitudes}

In this appendix, we outline how to evaluate the overlap integral,
$\textsf{O}_{a^{(*)}b^{(*)}c^{(*)}d^{(*)}}=F\int\left(\boldsymbol{\Phi}_{a}^{(*)}\cdot\boldsymbol{\Phi}_{b}^{(*)}\right)\left(\boldsymbol{\Phi}_{c}^{(*)}\cdot\boldsymbol{\Phi}_{d}^{(*)}\right)\text{d}\phi$.
The sum of three $\textsf{O}$ integrals gives the four-wave mixing
process amplitude $\textsf{U}=\textsf{O}_{i^{*}ps^{*}p'}+\textsf{O}_{i^{*}p's^{*}p}+\textsf{O}_{pp's^{*}i^{*}}$
for the set of modes in fields $v=p,p',s,i$. We label the mode indices
for field $v$ with the corresponding subscript, e.g., $s_{p},$ $L_{i}$,
or $S_{s}.$ As explained in the main body of the paper, a complex
conjugate on the field $v$ in the subscript of $\textsf{O}$ indicates
the corresponding mode should be conjugated, i.e., $\boldsymbol{\Phi}_{v}^{*}$.
In all three $\textsf{O}$ integrals, two of the indices are conjugated.

\subsubsection{Process Amplitudes for Definite Spin and Orbital Angular Momentum States}

We start by considering FWM processes in which all four beams are
in orbital and spin angular momentum eigenstate modes, $\mathrm{\mathbf{Y}}^{[l,s]}=e^{il\phi}\boldsymbol{\sigma}^{[s]}$.
The process amplitude, $\textsf{U}^{(\mathrm{Y})}$ is calculated
from the coupled amplitude equation for the signal Eq. (\ref{processamplitudes}).
The terms comprising the FWM process amplitude are formed by the product
of four $\mathrm{\mathbf{Y}}$
modes. We begin with half of this product: the dot product of two general $\mathrm{\mathbf{Y}}$
modes. For fields $a$ and $b$ this is 
\begin{equation}
\boldsymbol{\mathrm{Y}}_{a}^{(*)}\cdot\boldsymbol{\mathrm{Y}}_{b}^{(*)}=\boldsymbol{\sigma}^{[\pm s_{a}]}\cdot\boldsymbol{\sigma}^{[\pm s_{b}]}\mathrm{e}^{i(\pm l_{a}\pm l_{b})\phi}.\label{eq:inner_product}
\end{equation}
Here, $(*)$ indicates the optional presence of the complex conjugate
on $\boldsymbol{\Phi}_{v}$, in which case each mode index of the
$v$ field (e.g., $s_{v}$) is preceded by the bottom symbol of $\pm$
or $\mp$. Using $\boldsymbol{\sigma}^{[s]}=\frac{\left(\mathbf{x}+is\mathbf{y}\right)}{\sqrt{2}}$
with $s=\pm1$, it holds that $\boldsymbol{\sigma}^{[s_{i}]}\cdot\boldsymbol{\sigma}^{[s_{j}]}=\delta_{s_{i},-s_{j}}.$
The full argument of one overlap integral is then 
\begin{equation}
\begin{split}\textsf{O}_{a^{(*)}b^{(*)}c^{(*)}d^{(*)}}= & F\int\left(\boldsymbol{\mathrm{Y}}_{a}^{(*)}\cdot\boldsymbol{\mathrm{Y}}_{b}^{(*)}\right)\left(\boldsymbol{\mathrm{Y}}_{c}^{(*)}\cdot\boldsymbol{\mathrm{Y}}_{d}^{(*)}\right)\text{d}\phi\\
= & \delta_{\pm s_{a},\mp s_{b}}\delta_{\pm s_{c},\mp s_{d}}F\int\mathrm{e}^{i(\pm l_{a}\pm l_{b}\pm l_{c}\pm l_{d})\phi}\text{d}\phi\\
= & 2\pi F\delta_{\pm s_{a},\mp s_{b}}\delta_{\pm s_{c},\mp s_{d}}\delta_{0,(\pm l_{a}\pm l_{b}\pm l_{c}\pm l_{d}),}
\end{split}
\label{eq:Y_Overlap}
\end{equation}
where in the last line we use $\int_{0}^{2\pi}\exp(iq\phi)\text{d}\phi=2\pi\delta_{0,q}$.
Using this result, the three $\textsf{O}$ integrals composing $\textsf{U}^{(\mathrm{Y})}$
are 
\begin{flalign}
\textsf{O}_{i^{*}ps^{*}p'}= & 2\pi F\delta_{s_{i},s_{p}}\delta_{s_{s},s_{p'}}\delta_{l_{p}+l_{p'},l_{i}+l_{s}},\label{AI1}\\
\textsf{O}_{i^{*}p's^{*}p}= & 2\pi F\delta_{s_{i},s_{p'}}\delta_{s_{s},s_{p}}\delta_{l_{p}+l_{p'},l_{i}+l_{s}},\label{AI2}\\
\textsf{O}_{pp's^{*}i^{*}}= & 2\pi F\delta_{s_{p},-s_{p'}}\delta_{-s_{s},s_{i}}\delta_{l_{p}+l_{p'},l_{i}+l_{s}}.
\label{AI3}
\end{flalign}
The total process amplitude is then the summation of Eqs. (\ref{AI1}-\ref{AI3})
\begin{equation}
\begin{split}\textsf{U}^{(\mathrm{Y})}= & 2\pi F\left(\delta_{s_{i},s_{p}}\delta_{s_{s},s_{p'}}+\delta_{s_{i},s_{p'}}\delta_{s_{s},s_{p}}+\delta_{s_{p},-s_{p'}}\delta_{-s_{s},s_{i}}\right)\delta_{l_{p}+l_{p'},l_{i}+l_{s}}\\
= & 4\pi F\delta_{s_{i}+s_{s},s_{p}+s_{p'}}\delta_{l_{i}+l_{s},l_{p}+l_{p'}},
\end{split}
\label{eq:U_Y}
\end{equation}
where the last line uses an identity that we will introduce in the
next section. Each Kronecker delta provides a selection rule: $s_{i}+s_{s}=s_{p}+s_{p'}$
and $l_{i}+l_{s}=l_{p}+l_{p'}$, or more succinctly, $\Delta s=0$
and $\Delta l=0$. In summary, the SAM and OAM are independently conserved
in the FWM process.

\subsubsection{Process Amplitudes for the CV Modes}

We now calculate the process amplitude $\textsf{U}^{(\mathrm{CV})}$
for a four-wave mixing process where all the beams are in CV modes.
A CV mode consists of a superposition of two $\mathbf{Y}$ modes,
as in Eq. (\ref{eq:CV_modes}): 
\begin{equation}
\mathbf{CV}^{[L,S]}=\frac{1}{\sqrt{2}}\left(\mathrm{e}^{i\frac{\pi}{4}\left(S-1\right)}\mathrm{\mathbf{Y}}^{[L,S]}+\mathrm{e}^{-i\frac{\pi}{4}\left(S-1\right)}\mathrm{\mathbf{Y}}^{[-L,-S]}\right).\label{eq:CV_recap}
\end{equation}
Since these modes are real, we drop the complex conjugates in $\textsf{O}.$
The mode overlap integral, Eq. (\ref{eq:modeoverlap}), contains two
inner products of two CV modes each. When expanded using Eq. (\ref{eq:CV_modes}),
$\textsf{O}$ contains 16 terms, each containing a $\mathbf{Y}$ mode
factor from all four fields, $v=a,b,c,d$. From Eq. (\ref{eq:CV_recap}),
each of these four $\mathbf{Y}$ mode factors will either be of the
form $\mathrm{\mathbf{Y}}^{[+L_{v},+S_{v}]}$ or $\mathrm{\mathbf{Y}}^{[-L_{v},-S_{v}]}$.
Consequently, we represent these 16 terms as a sum indexed by the
sign of the $S$ and $L$ values, as represented by $n_{v}=\pm1$:
\begin{eqnarray}
\textsf{O}_{abcd} & = & F\int\left(\mathrm{\mathbf{CV}}^{[L_{a},S_{a}]}\cdot\mathrm{\mathbf{CV}}^{[L_{b},S_{b}]}\right)\left(\mathrm{\mathbf{CV}}^{[L_{c},S_{c}]}\cdot\mathrm{\mathbf{CV}}^{[L_{d},S_{d}]}\right)\text{d}\phi\nonumber \\
 & = & \frac{F}{4}\sum_{n_{a}..n_{d}=\pm1}\mathrm{e}^{i\frac{\pi}{4}\left(n_{a}S_{a}+n_{b}S_{b}+n_{c}S_{c}+n_{d}S_{d}-K\right)}\nonumber \\
 & \times & \int\left(\mathrm{\mathbf{Y}}^{[n_{a}L_{a},n_{a}S_{a}]}\cdot\mathrm{\mathbf{Y}}^{[n_{b}L_{b},n_{b}S_{b}]}\right)\left(\mathrm{\mathbf{Y}}^{[n_{c}L_{c},n_{c}S_{c}]}\cdot\mathrm{\mathbf{Y}}^{[n_{d}L_{d},n_{d}S_{d}]}\right)\text{d}\phi\nonumber \\
 & = & \frac{\pi F}{2}\sum_{n_{a}..n_{d}=\pm1}\mathrm{e}^{i\frac{\pi}{4}\left(n_{a}S_{a}+n_{b}S_{b}+n_{c}S_{c}+n_{d}S_{d}-K\right)}\nonumber \\
 & \times & \delta_{n_{a}S_{a},-n_{b}S_{b}}\delta_{n_{c}S_{c},-n_{d}S_{d}}\delta_{0,n_{a}L+n_{b}L_{b}+n_{c}L_{c}+n_{d}L_{d}}\nonumber \\
 & = & \frac{\pi F}{2}\sum_{n_{a}..n_{d}=\pm1}\mathrm{e}^{-i\frac{\pi}{4}K}\delta_{n_{a}S_{a},-n_{b}S_{b}}\delta_{n_{c}S_{c},-n_{d}S_{d}}\nonumber \\
 & \times & \delta_{0,n_{a}L+n_{b}L_{b}+n_{c}L_{c}+n_{d}L_{d}},\label{eq:eqarrayex}
\end{eqnarray}
where the azimuthal integral was evaluated according to the last section.
Each term differs by a constant term in the exponential, $K=n_{a}+n_{b}+n_{c}+n_{d}$.
In the final expression, we applied the $S$ Kronecker deltas to the
exponential factor. 

We will now simplify this expression by pairing the sixteen terms
with their complex conjugates (i.e., the term with a particular set
of $n_{a},n_{b},n_{c},n_{d}$ values is conjugate to the $-n_{a},-n_{b},-n_{c},-n_{d}$
term). Combining each conjugate pair, we find the general form for
each of the eight resulting terms: 
\begin{eqnarray}
G_{n_{a}n_{b}n_{c}n_{d}}=\pi F\cos\left(\frac{\pi}{4}K\right)\delta_{n_{a}S_{a},-n_{b}S_{b}}\delta_{n_{c}S_{c},-n_{d}S_{d}}\nonumber \\
\times\delta_{0,n_{a}L+n_{b}L_{b}+n_{c}L_{c}+n_{d}L_{d}}.
\end{eqnarray}
With this definition, $\textsf{O}_{abcd}=G_{++++}+G_{+++-}+G_{++-+}+G_{++--}+G_{+-++}+G_{+-+-}+G_{+--+}+G_{-+++}.$
Each of the eight terms has either zero, one, or two $n_{v}$ parameters
with $n_{v}=-1$ (the other variations of $\{n_{v}\}$ were covered
by the eight complex conjugate terms). Consequently, $K=4,2,$ and
0 for these cases, respectively. If $K=2$ then the cos term is
$0$, which immediately eliminates half the remaining terms, leaving
$\textsf{O}_{abcd}=G_{++++}+G_{++--}+G_{+-+-}+G_{+--+}$. If $K=0$,
then the cos factor equals $1$ and if $K=4$ it equals $-1$.
Applying this to each term we find 
\begin{align}
\textsf{O}_{abcd} & =\pi F\\
\times & \Biggl[-\delta_{S_{a},-S_{b}}\delta_{S_{c},-S_{d}}\delta_{0,L_{a}+L_{b}+L_{c}+L_{d}}\nonumber \\
+ & \delta_{S_{a},-S_{b}}\delta_{S_{c},-S_{d}}\delta_{0,L_{a}+L_{b}-L_{c}-L_{d}}\nonumber \\
+ & \delta_{S_{a},S_{b}}\delta_{S_{c},S_{d}}\delta_{0,L_{a}-L_{b}+L_{c}-L_{d}}\nonumber \\
+ & \delta_{S_{a},S_{b}}\delta_{S_{c},S_{d}}\delta_{0,L_{a}-L_{b}-L_{c}+L_{d}}\Biggr],
\end{align}
where we have used $\delta_{n_{j}S_{j},n_{k}S_{k}}=\delta_{-n_{j}S_{j},-n_{k}S_{k}}$
to ensure that the first argument in all the $S$ Kronecker deltas
is positive.

\begin{table*}[htb!]
\centering \captionsetup{justification=centering} %
\caption{Selection Rules for the CV Modes.}
\label{CVselectrule2} 
\begin{tabular}{|c|c|c|c|}
\hline 
Rule  & $L$  & $S$  & Amplitude\tabularnewline
\hline 
1\&2  & $L_{p}+L_{p'}=\pm\left(L_{s}+L_{i}\right)$  & $S_{p}+S_{p'}=\pm\left(S_{s}+S_{i}\right)$  & $\pm2\pi F$\tabularnewline
\hline 
3\&4  & $L_{p}-L_{p'}=\pm\left(L_{s}-L_{i}\right)$  & $S_{p}-S_{p'}=\pm\left(S_{s}-S_{i}\right)$  & $2\pi F$\tabularnewline
\hline 
\end{tabular}
\end{table*}

Now we find the three overlap integrals that compose $\textsf{U}^{(\mathrm{CV})}$:

\begin{align}
\textsf{O}_{i^{*}ps^{*}p'} & =\pi F\nonumber \\
\times & \Biggl[-\delta_{S_{i},-S_{p}}\delta_{S_{s},-S_{p'}}\delta_{0,L_{i}+L_{p}+L_{s}+L_{p'}}\nonumber \\
+ & \delta_{S_{i},-S_{p}}\delta_{S_{s},-S_{p'}}\delta_{0,L_{i}+L_{p}-L_{s}-L_{p'}}\nonumber \\
+ & \delta_{S_{i},S_{p}}\delta_{S_{s},S_{p'}}\delta_{0,L_{i}-L_{p}+L_{s}-L_{p'}}\nonumber \\
+ & \delta_{S_{i},S_{p}}\delta_{S_{s},S_{p'}}\delta_{0,L_{i}-L_{p}-L_{s}+L_{p'}}\Biggr],
\end{align}
\begin{align}
\textsf{O}_{i^{*}p's^{*}p} & =\pi F\nonumber \\
\times & \Biggl[-\delta_{S_{i},-S_{p'}}\delta_{S_{s},-S_{p}}\delta_{0,L_{i}+L_{p'}+L_{s}+L_{p}}\nonumber \\
+ & \delta_{S_{i},-S_{p'}}\delta_{S_{s},-S_{p}}\delta_{0,L_{i}+L_{p'}-L_{s}-L_{p}}\nonumber \\
+ & \delta_{S_{i},S_{p'}}\delta_{S_{s},S_{p}}\delta_{0,L_{i}-L_{p'}+L_{s}-L_{p}}\nonumber \\
+ & \delta_{S_{i},S_{p'}}\delta_{S_{s},S_{p}}\delta_{0,L_{i}-L_{p'}-L_{s}+L_{p}}\Biggr],
\end{align}
\begin{align}
\textsf{O}_{pp's^{*}i^{*}} & =\pi F\nonumber \\
\times & \Biggl[-\delta_{S_{p},-S_{p'}}\delta_{S_{s},-S_{i}}\delta_{0,L_{p}+L_{p'}+L_{s}+L_{i}}\nonumber \\
+ & \delta_{S_{p},-S_{p'}}\delta_{S_{s},-S_{i}}\delta_{0,L_{p}+L_{p'}-L_{s}-L_{i}}\nonumber \\
+ & \delta_{S_{p},S_{p'}}\delta_{S_{s},S_{i}}\delta_{0,L_{p}-L_{p'}+L_{s}-L_{i}}\nonumber \\
+ & \delta_{S_{p},S_{p'}}\delta_{S_{s},S_{i}}\delta_{0,L_{p}-L_{p'}-L_{s}+L_{i}}\Biggr].
\end{align}
Notice how the same four $L$ Kronecker deltas appear in each $\textsf{O}$,
albeit in a different order each time. In the $\textsf{U}^{(\mathrm{CV})}$
sum, we now group these $L$ Kronecker delta terms together:

\begin{align}
\textsf{U}^{(\mathrm{CV})} & =\textsf{O}_{i^{*}ps^{*}p'}+\textsf{O}_{i^{*}p's^{*}p}+\textsf{O}_{pp's^{*}i^{*}}\nonumber \\
 & =\pi F\nonumber \\
\times & \Biggl[-\left(\delta_{S_{i},-S_{p}}\delta_{S_{s},-S_{p'}}+\delta_{S_{i},-S_{p'}}\delta_{S_{s},-S_{p}}+\delta_{S_{p},-S_{p'}}\delta_{S_{s},-S_{i}}\right)\nonumber \\
\times & \delta_{0,L_{i}+L_{p}+L_{s}+L_{p'}}\nonumber \\
+ & \left(\delta_{S_{i},-S_{p}}\delta_{S_{s},-S_{p'}}+\delta_{S_{i},S_{p'}}\delta_{S_{s},S_{p}}+\delta_{S_{p},S_{p'}}\delta_{S_{s},S_{i}}\right)\delta_{0,L_{i}+L_{p}-L_{s}-L_{p'}}\nonumber \\
+ & \left(\delta_{S_{i},S_{p}}\delta_{S_{s},S_{p'}}+\delta_{S_{i},S_{p'}}\delta_{S_{s},S_{p}}+\delta_{S_{p},-S_{p'}}\delta_{S_{s},-S_{i}}\right)\delta_{0,L_{i}-L_{p}+L_{s}-L_{p'}}\nonumber \\
+ & \left(\delta_{S_{i},S_{p}}\delta_{S_{s},S_{p'}}+\delta_{S_{i},-S_{p'}}\delta_{S_{s},-S_{p}}+\delta_{S_{p},S_{p'}}\delta_{S_{s},S_{i}}\right)\delta_{0,L_{i}-L_{p}-L_{s}+L_{p'}}\Biggr].
\end{align}

The following identity will be used to considerably simplify the expression
for $\textsf{U}^{(\mathrm{CV})},$ 
\begin{equation}
\begin{split}\delta_{S_{a},q_{1}S_{c}}\delta_{S_{b},q_{1}S_{d}}+\delta_{S_{a},q_{2}S_{d}}\delta_{S_{b},q_{2}S_{c}}+\delta_{S_{a},-q_{1}q_{2}S_{b}}\delta_{S_{c},-q_{1}q_{2}S_{d}}\\
=2\delta_{S_{b}+q_{1}q_{2}S_{a},q_{2}S_{c}+q_{1}S_{d}},
\end{split}
\end{equation}
where $q_{k}=\pm1.$ While we omit a detailed proof, the identity's
validity can be understood by considering the three possible values
of either of the arguments of the Kronecker delta on the RHS, -2,0,
or 2. For both delta arguments to be 2 or both arguments to be -2,
all the terms in both arguments must be equal. The first two terms
on the LHS sum to two if all the terms on the RHS are equal, thereby
covering these cases. The third term and one of the other first two
terms on the LHS cover the remaining case, in which both RHS delta
arguments are 0.

Applying this identity to $\textsf{U}^{(\mathrm{CV})}$ four times
and reordering the arguments of the Kronecker deltas we find: 
\begin{align}
\textsf{U}^{(\mathrm{CV})} & =2\pi F\nonumber \\
\times & \Biggl[-\delta_{S_{s}+S_{i},-(S_{p}+S_{p'})}\delta_{L_{s}+L_{i},-(L_{p}+L_{p'})}\nonumber \\
+ & \delta_{S_{s}-S_{i},S_{p}-S_{p'}}\delta_{L_{s}-L_{i},L_{p}-L_{p'}}\nonumber \\
+ & \delta_{S_{s}+S_{i},S_{p}+S_{p'}}\delta_{L_{s}+L_{i},L_{p}+L_{p'}}\label{eq:U_CV}\\
+ & \delta_{S_{s}-S_{i},-(S_{p}-S_{p'})}\delta_{L_{s}-L_{i},-(L_{p}-L_{p'})}\Biggr].\nonumber 
\end{align} These four terms correspond to four selection rules, each of which
has an $L$ sub-rule and $S$ sub-rule. These are summarized in Table~\ref{CVselectrule2}.

In Table \ref{CVselectrule2}, either the top or bottom symbol of $\pm$ and $\mp$ should be taken consistently throughout each rule.
More than one rule can hold true simultaneously and their amplitudes
should be added to find $\textsf{U}$, the total process amplitude. 

\begin{figure}
\protect\includegraphics{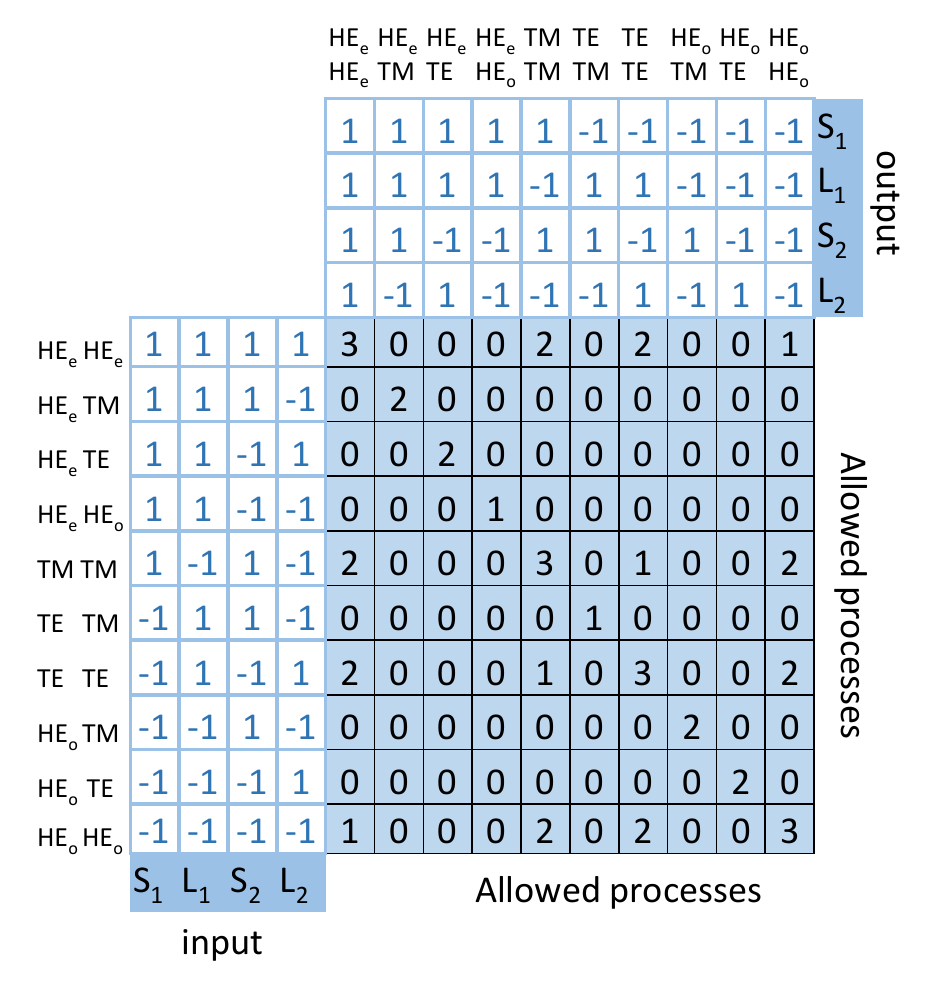}\protect\caption{Process amplitudes $\textsf{U}$ for four-wave mixing between CV modes.
All four modes have the same value of $|L|=|l|$. Here, we take $|l|=1$
but the amplitudes for $|l|>1$ will be identical to these. The process
amplitudes in the table are calculated from Eq. (\ref{eq:U_CV}) and
are normalized as $\textsf{U}/2\pi F$. \label{processamplitudes}}
\end{figure}

In Fig. \ref{processamplitudes}, we compile the amplitudes ($\times2\pi F$)
for all distinguishable processes. We consider only the simplest case,
one in which all four fields are chosen from the set of four CV modes
defined by having the same value of $|L|=|l|$. As an example of this,
we will identify the relevant modes for $|L|=1$ (i.e., TM, TE, etc.),
but the results hold for all $|L|$. The labels $p$ and $p'$ are
interchangeable without changing the physical process, likewise for
output fields $s$ and $i$. In accordance with this, we simply label
the input and output mode indices by subscript 1 and 2. With this
in mind, in our scenario, there are ten physically distinct mode combinations
for the pump modes and the same for the output modes. (In combinatorial
multiset notation, $\left(\left(\begin{smallmatrix}n\\
k
\end{smallmatrix}\right)\right)=10$ for $n=4$ modes and $k=2$ input/output fields.) It follows that
there are $10\times10=100$ potential distinct FWM processes in total.

\subsubsection{Process Amplitudes for the Total Angular Momentum modes}

We can use the results of the previous two sections to find selection
rules for the TAM mode basis, that is, the modes with definite $j$
that are simultaneously fiber eigenmodes. We categorize the potential
processes into three cases, which we treat separately in the following
subsections. In each, we find $\textsf{U}^{(\mathrm{TAM})}$, the
process amplitude. The results are summarized in Table \ref{allowedTAM}.

\paragraph{Case 1: Interconversion of $j=0$ modes}

First we consider interconversion between the two $j=0$ modes, $\mathbf{Z}^{[-0]}$
and $\mathbf{Z}^{[+0]}$ (i.e., the TE and TM modes, respectively).
All the potential processes will trivially conserve $j$ since it
is equal to 0 for every involved mode. Since these are $\mathrm{CV}$
modes, these cases will be governed by the corresponding selection
rules and amplitudes from Eq. (\ref{eq:U_CV}). Like in the previous
section, the only disallowed processes are ones in which there is
a single unpaired mode among the four modes in the process, e.g.,
$\mathrm{TM}+\mathrm{TE}\rightarrow\mathrm{TM}+\mathrm{TM}$. Note
that this process would be allowed by total angular momentum conservation,
so this principle by itself is not the sole selection rule for the
TAM basis. If all the modes are alike $\textsf{U}^{(\mathrm{TAM})}=6\pi F$,
otherwise $\textsf{U}^{(\mathrm{TAM})}=2\pi F$.

\paragraph{Case 2: Interconversion of $j=$$\pm2$ modes}

Similarly, we can consider conversion between the $j=\pm2$ modes
in the TAM basis. These are the $\mathbf{Z}^{[2]}=\mathrm{\mathbf{Y}}^{[1,1]}$
and $\mathbf{Z}^{[-2]}=\mathrm{\mathbf{Y}}^{[-1,-1]}$ modes. The
relevant selection rules, given by Eq. (\ref{eq:U_Y}), are simply
those that separately conserve SAM $s$ and OAM $l$, $s_{i}+s_{s}=s_{p}+s_{p'}$
and $l_{i}+l_{s}=l_{p}+l_{p'}$. Since in each beam $s=l$, one type
of angular momentum conservation is automatically accompanied by the
other.
Additionally, the total angular momentum will also be trivially conserved
since $j=s+l$. Conversely, the only way to conserve TAM for these
modes is for SAM and OAM to be conserved.

\paragraph{Case 3: Conversion between $j=0$ and $j=\pm2$ modes}

Lastly, we consider interconversion between the $j=\pm2$ modes and
the $j=0$ modes. We describe the two $j=0$ modes $\mathbf{Z}^{[\pm0]}$
with a single expression that depends on a single parameter, $m=\pm1$:
\begin{eqnarray}
\mathbf{Z}^{[m0]}=\mathbf{CV}^{[-m,m]}=\frac{1}{\sqrt{2}}\mathrm{e}^{i\frac{\pi}{4}\left(m-1\right)}\left(\mathrm{\mathbf{Y}}^{[1,-1]}+m\mathrm{\mathbf{Y}}^{[-1,1]}\right).\label{eq:TAM_single}
\end{eqnarray}
Since each overlap integral $\textsf{O}$ is linear in a given beam's
mode, if that beam is in a $\mathbf{Z}$ mode the number of terms
in $\textsf{U}^{(\mathrm{TAM})}$ will double. Each term will correspond
to a process amplitude $\textsf{U}^{(\mathrm{Y})}$for the $\mathrm{\mathbf{Y}}$
modes in Eq. (\ref{eq:TAM_single}). (We will show how this works
below). We will derive the selection rules for the TAM modes using
the selection rules of the constituent $\mathrm{\mathbf{Y}}$ modes.
Key will be the fact that each field $q$ in a $\mathbf{Z}$ mode
will have anti-aligned SAM and OAM, $s_{q}=-l_{q},$ whereas they
will be aligned for each beam in a $\mathrm{\mathbf{Y}}$ mode, $s_{q}=l_{q}.$
We now separately consider the sub-cases in which one, two, or three
of the beams are in $\mathbf{Z}$ modes.

\subparagraph{Sub-case 3a: One beam in a $j=0$ mode.}

We start with the case of a sole beam in a $\mathbf{Z}$ mode, say
the idler beam. By linearity, the superposition of two $\mathrm{\mathbf{Y}}$
modes in the $\mathbf{Z}$ factor leads to two terms in the TAM process
amplitude: 
\begin{equation}
\textsf{U}^{(\mathrm{TAM})}=\frac{1}{\sqrt{2}}\mathrm{e}^{i\frac{\pi}{4}\left(m_{i}-1\right)}\left(\textsf{U}_{+spp'}^{(\mathrm{Y})}+m_{i}\textsf{U}_{-spp'}^{(\mathrm{Y})}\right).\label{eq:UTAM}
\end{equation}
Here, $\textsf{U}_{ispp'}^{(\mathrm{Y})}$ is the process amplitude
for four fields in $\mathrm{\mathbf{Y}}$ modes (i.e., Eq. (\ref{eq:U_Y})),
where if the field subscript $v=i,s,p,p'$ is replaced by $n_{v}=\pm1$
it indicates that field $v$ is in mode $\mathrm{\mathbf{Y}}^{[n_{v},-n_{v}]}$.
Each $\textsf{U}_{ispp'}^{(\mathrm{Y})}$ term is only nonzero if
it satisfies the standard $\mathrm{\mathbf{Y}}$ mode selection rules,
in which case $\textsf{U}_{ispp'}^{(\mathrm{Y})}=4\pi F,$ regardless
of which four $\mathrm{\mathbf{Y}}$ modes are involved. At least
one term must be an allowed process for $\textsf{U}^{(\mathrm{TAM})}$
to be nonzero. In the current case, the $\mathrm{\mathbf{Y}}$ rules
become $n_{i}+s_{s}=s_{p}+s_{p'}$ and $-n_{i}+l_{s}=l_{p}+l_{p'}$
for each of the two $\textsf{U}_{n_{i}spp'}^{(\mathrm{Y})}$ terms,
where $n_{i}=\pm1$ sets the term considered. However, since for all
the other fields $s_{j}=l_{j}$ (i.e., modes $j=i,s,p$, are in $\mathrm{\mathbf{Y}}^{[\pm n_{j},\pm n_{j}]}$),
we have $n_{i}+s_{s}=s_{p}+s_{p'}$ and $-n_{i}+s_{s}=s_{p}+s_{p'}$.
Subtracting the two expressions, we are left with $n_{i}=0,$ which
contradicts $n_{i}=\pm1$. Consequently, neither term is allowed and
$\textsf{U}^{(\mathrm{TAM})}$ is always zero in the current case.
In other words, a process involving a single $\mathbf{Z}$ mode is
not allowed.

\subparagraph{Sub-case 3b: Both input beams or both output beams in $j=0$ modes.}

We now move on to the case with two input beams in $\mathbf{Z}$ modes.
The reasoning will be similar but now we must also consider the relative
amplitudes of the terms, since they could cancel each other. Though
we shall not describe it explicitly, the case with two output beams
in $\mathbf{Z}$ modes follows the same selection rules. Now that
there are two $\mathbf{Z}$ modes, there are four terms in the process
amplitude: 
\begin{eqnarray}
\begin{aligned}\textsf{U}^{(\mathrm{TAM})}= & \frac{1}{2}\mathrm{e}^{i\frac{\pi}{4}\left(m_{i}+m_{s}-2\right)}\biggl(\textsf{U}_{++pp'}^{(\mathrm{Y})}+m_{i}\textsf{U}_{-+pp'}^{(\mathrm{Y})}\\
 & +m_{s}\textsf{U}_{+-pp'}^{(\mathrm{Y})}+m_{i}m_{s}\textsf{U}_{--pp'}^{(\mathrm{Y})}\biggr)\\
= & \frac{1}{2}\mathrm{e}^{i\frac{\pi}{4}\left(m_{i}+m_{s}-2\right)}\left(m_{i}\textsf{U}_{-+pp'}^{(\mathrm{Y})}+m_{s}\textsf{U}_{+-pp'}^{(\mathrm{Y})}\right)\\
= & \frac{1}{2}\mathrm{e}^{i\frac{\pi}{4}\left(m_{i}+m_{s}-2\right)}4\pi F\left(m_{i}+m_{s}\right),
\end{aligned}
\end{eqnarray}

where we shall now explain the steps between the lines. Analogous
to the last case, each term must satisfy $n_{i}+n_{s}=s_{p}+s_{p'}$
and $-n_{i}-n_{s}=s_{p}+s_{p'}$. Adding and subtracting these, we
are left with $n_{i}=-n_{s}$ and $s_{p}=-s_{p'}$. The latter must
be true for the whole process. The former sets which terms are non-zero.
The second line above is the remaining terms, the cross-terms (e.g.,
$n_{i}=1,n_{s}=-1$) between the $\mathbf{Z}$ mode superpositions.
The process amplitudes are equal, $\textsf{U}_{-+pp'}^{(\mathrm{Y})}=\textsf{U}_{+-pp'}^{(\mathrm{Y})}=4\pi F,$
giving the third line above. Consequently, in this sub-case, $\textsf{U}^{(\mathrm{TAM})}$
is non-zero only if $m_{i}=m_{s}$. In other words, the two $\mathbf{Z}$
modes must be the same (e.g., $m_{i}=m_{s}$) and the $\mathrm{\mathbf{Y}}$
modes must be opposite (e.g., $s_{p}=-s_{p'}$). The total process
amplitude is $\textsf{U}^{(\mathrm{TAM})}=\pm4\pi F\mathrm{e}^{i\frac{\pi}{4}\left(\pm2-2\right)}=4\pi F$.

\subparagraph{Sub-case 3c: One input beam and one output beam in $j=0$ modes.}

The next case, where one input beam and one output beam are in a $\mathbf{Z}$
mode, follows similar reasoning to the last. And so, $n_{i}+s_{s}=n_{p}+s_{p'}$
and $-n_{i}+s_{s}=-n_{p}+s_{p'}$ must be satisfied for a term to
be nonzero. This leads to $s_{s}=s_{p'}$ and $n_{i}=n_{p'}$. With
this, $\textsf{U}^{(\mathrm{TAM})}=\frac{1}{2}\mathrm{e}^{i\frac{\pi}{4}\left(m_{i}+m_{s}-2\right)}4\pi F\left(1+m_{i}m_{s}\right)$.
Thus, $\textsf{U}^{(\mathrm{TAM})}$ is nonzero only if $m_{i}=m_{s}$,
which means the input beam is in the same $\mathbf{Z}$ mode to the
pump and the remaining input and pump beams are in the same $\mathrm{\mathbf{Y}}$
mode (e.g., $s_{s}=s_{p'}$). Again, the total process amplitude is
$\textsf{U}^{(\mathrm{TAM})}=4\pi F$.

\subparagraph{Sub-case 3d: Three beams in $j=0$ modes.}

\begin{table*}[ht]
\centering \captionsetup{justification=centering} %
\caption{Summary of Allowed FWM Processes Between TAM modes.}
\label{allowedTAM} 
\begin{tabular}{|c|c|c|c|}
\hline 
Case  & Process Type  & Allowed Processes  & Amplitude, $\textsf{U}^{(\mathrm{TAM})}$\tabularnewline
\hline 
1  & $\mathbf{Z}^{[w0]}+\mathbf{Z}^{[q0]}\rightarrow\mathbf{Z}^{[q0]}+\mathbf{Z}^{[q0]}$  & $w=\pm,q=\pm$  & $\begin{cases}
2\pi F & \textrm{for }w\neq q\\
6\pi F & \textrm{for }w=q
\end{cases}$\tabularnewline
\hline 
2  & $\mathrm{\mathbf{Y}}^{[s_{p},s_{p}]}+\mathrm{\mathbf{Y}}^{[s_{p'},s_{p'}]}\rightarrow\mathrm{\mathbf{Y}}^{[s_{i},s_{i}]}+\mathrm{\mathbf{Y}}^{[s_{s},s_{s}]}$  & $s_{p}+s_{p'}=s_{i}+s_{s}$  & $4\pi F$\tabularnewline
\hline 
3b  & $\mathrm{\mathbf{Z}}^{[w0]}+\mathrm{\mathbf{Z}}^{[w0]}\rightarrow\mathrm{\mathbf{Y}}^{[q,q]}+\mathrm{\mathbf{Y}}^{[-q,-q]}$  & $w=\pm,q=\pm1$  & $4\pi F$\tabularnewline
\hline 
3b  & $\mathrm{\mathbf{Y}}^{[q,q]}+\mathrm{\mathbf{Y}}^{[-q,-q]}\rightarrow\mathrm{\mathbf{Z}}^{[w0]}+\mathrm{\mathbf{Z}}^{[w0]}$  & $w=\pm,q=\pm1$  & $4\pi F$\tabularnewline
\hline 
3c  & $\mathrm{\mathbf{Z}}^{[w0]}+\mathrm{\mathbf{Y}}^{[q,q]}\rightarrow\mathrm{\mathbf{Z}}^{[w0]}+\mathrm{\mathbf{Y}}^{[q,q]}$  & $w=\pm,q=\pm1$  & $4\pi F$\tabularnewline
\hline 
\end{tabular}
\end{table*}
The last case, three of the four beams in $\mathbf{Z}$ modes, is
relatively simple. There will be six terms in $\textsf{U}^{(\mathrm{TAM})}$,
each of which must satisfy $n_{i}+n_{s}=n_{p}+s_{p'}$ and $-n_{i}-n_{s}=-n_{p}+s_{p'}$.
Combining these equations leads to the contradiction, $s_{p'}=0$.
Consequently, no such process is possible.

All the allowed processes and their relative amplitudes are listed
in the following Table \ref{allowedTAM}.

\newpage
\bibliography{PaperCitations}

\end{document}